\documentclass[12pt,letterpaper]{article}
\pdfoutput=1
\usepackage{youngtab}

\usepackage{bbm}
\usepackage{amscd,amssymb,epsf}

\usepackage{textcomp}

\usepackage[latin1]{inputenc}
\usepackage{amsfonts}
\usepackage{amssymb}
\usepackage{amsmath}
\usepackage{dsfont}
\usepackage{hyperref}
\usepackage{slashed}
\usepackage{empheq}
\usepackage{multirow}
\usepackage{bbm}
\usepackage{fancyhdr}
\hypersetup{
            pdfstartview=FitH,
            pdfpagemode=None,
            pdfborder={0.75 0.75 0.75},
            linkbordercolor={0.75 0.75 1}
            }
\newcommand{\be}{\begin{equation}}
\newcommand{\ee}{\end{equation}}
\newcommand{\beq}{\begin{equation}}
\newcommand{\eeq}{\end{equation}}
\newcommand{\bea}{\begin{eqnarray}}
\newcommand{\eea}{\end{eqnarray}}

\numberwithin{equation}{section}
\begin{document}
\title{{\bf The Spectrum of Excitations of Holographic Wilson
Loops}}
\author{Alberto Faraggi and Leopoldo A. Pando Zayas \\\\
{\small\emph{Michigan Center for Theoretical Physics}} \\ {\small\emph{Randall Laboratory of Physics, The University of Michigan}} \\ {\small\emph{Ann Arbor, MI 48109-1040}} \\\\ {\small \texttt{faraggi@umich.edu, lpandoz@umich.edu}}}
\date{}
\maketitle
\thispagestyle{fancy}
\rhead{MCTP-11-02}
\cfoot{}
\renewcommand{\headrulewidth}{0pt}
\begin{abstract}
In the holographic framework, a half BPS Wilson loop in ${\cal N}=4$
supersymmetric Yang-Mills in the fundamental, symmetric or antisymmetric representation of $SU(N)$,
is best described by a fundamental string, a D3 brane or a D5 brane with fluxes in their worldvolumes, respectively. We
derive the spectrum of excitations of such D3 brane in $AdS_5\times
S^5$ explicitly, considering its action in both the bosonic and the
fermionic sectors, and demonstrate that it is organized
according to short multiplets of the supergroup $OSp(4^*|4)$. We also show that the modes of the fundamental string form an ultra-short multiplet of this supergroup. In the case of the D5 brane the bosonic spectrum is only partially known but we argue that it also fills representations of $OSp(4^*|4)$. This way we provide a step towards a unifying picture for the description of holographic excitations of the circular and straight supersymmetric Wilson loops in arbitrary representations.
\end{abstract}
\newpage
\tableofcontents
\newpage
\section{Introduction}
Wilson loops are important gauge invariant operators in gauge theories. It is possible to reformulate the theory in terms of these nonlocal operators and they also serve as useful order parameters. In the context of the AdS/CFT correspondence Wilson loops were first formulated by Maldacena \cite{Maldacena:1998im} and Rey-Yee \cite{Rey:1998ik}. The prescription was to identify the expectation value of Wilson loops with the action of a fundamental string in the dual supergravity background; Rey-Yee already mentioned the relevance of D-branes with worldvolume fluxes as potential decorating parameters of the Wilson loop.

A particularly important role is played by supersymmetric Wilson loops, most notably, the circular Wilson loop \cite{Drukker:1999zq}. Its expectation value was conjectured to be computed exactly via a Gaussian matrix model in \cite{Drukker:2000rr,Erickson:2000af}, with a later rigorous proof appearing in \cite{Pestun:2007rz}.

A nontrivial decoration of the circular Wilson loop is the representation of the gauge group. We have by now a good understanding of the fundamental, symmetric and the antisymmetric representations from the holographic point of view \cite{Drukker:1999zq,Drukker:2005kx} and its stringy origin \cite{Gomis:2006im,Gomis:2006sb}. The work of \cite{Drukker:2005kx} focused on the D3 which is dual to the Wilson loop in the symmetric representation. By analogy with the giant graviton argument, Yamaguchi developed the case of a D5 brane wrapping $S^4$ in $AdS_5\times S^5$ and identified it with the description of a Wilson loop in the antisymmetric representation \cite{Yamaguchi:2006tq}. Interestingly, using the Gaussian matrix model it was possible to confirm the finer structure of the representations  \cite{Yamaguchi:2006tq,Hartnoll:2006is,Yamaguchi:2007ps,Giombi:2006de}.

There is a very interesting characterization depending on the value of $k$, the number of boxes in the Young tableau of $SU(N)$, relative to $N$ in the large $N$ limit. When $k$ is order one, that is, $k\ll N$, the Wilson loop is effectively described, on the gravity side, by $k$ fundamental strings. For $k/N$ fixed, the most fitting holographic description is that of probe branes with fluxes. Finally, there could also be Wilson loops in representations with $k\sim N^2$; these are almost square Young tableaux. In this case the probe approximation is no longer valid and a fully backreacted supergravity background must be constructed. Such construction has been carried out in some simple cases in \cite{Yamaguchi:2006te,Lunin:2006xr,D'Hoker:2007fq} and further refined in \cite{Okuda:2008px,Gomis:2008qa}.

Of paramount importance is the computation of quantum corrections to the given expectations values. In a sense, this is the act of taking the AdS/CFT correspondence beyond the comparison of classical ground state configurations. This is  the equivalent to high-precision spectroscopy, that is,  an analysis of the quantum corrections is the way to validate our assumptions about the correspondence. Indeed, corrections to the circular Wilson loop dual to the fundamental string have been computed in various works \cite{Forste1999,Drukker:2000ep,Sakaguchi:2007ea,Kruczenski:2008zk}. A prescription for computing correlators of Wilson loops with chiral primaries and with another Wilson loop was developed in the early stages of the AdS/CFT correspondence \cite{Berenstein:1998ij}. This prescription was beautifully applied in the case of symmetric and antisymmetric Wilson loops with chiral primaries and was shown to coincide with the calculation from the matrix model in \cite{Giombi:2006de}.

In this paper our goal is to go beyond the ``ground state'' analysis of Wilson loops and study the excitations on the gravity side. We study the holographic description of the half BPS Wilson loop in the symmetric representation, that is, a D3 brane configuration in $AdS_5\times S^5$ whose worldvolume is $AdS_2\times S^2$. We explicitly compute the bosonic and fermionic fluctuations starting from the action for such a D3 brane as worked out by Martucci and collaborators \cite{Marolf:2003ye,Marolf:2003vf,Martucci:2005rb,Martucci:2005ht,Martucci:2006ij}. After this explicit calculation we fit the spectrum of excitations into short multiplets of $OSp(4^*|4)$. Emboldened by our success in the explicit case of the D3 brane, we go on and fit the excitations of the fundamental string and the D5 brane found in the literature into short multiplets of  $OSp(4^*|4)$. Thus, using mostly its symmetries, we present the spectrum of excitations of the holographic description of half BPS Wilson loops.

According to the construction in \cite{Gomis:2006im}, a Wilson loop in an arbitrary representation of $SU(N)$ has a holographic description in terms of coincident D3 branes or, alternatively, coincident D5 branes. A complete analysis of the spectrum then requires dealing with the non-Abelian nature of the corresponding low energy action. The case of a single D3 brane corresponds to a Young tableau with one row.

The paper is organized as follows. In section \ref{Sec:Class-config} we review the classical D3 brane configuration. Section \ref{Sec:D3} presents an explicit computation of the spectrum of excitations, both in the bosonic and fermionic sectors. Section \ref{Supersymmetry} contains a discussion of the supersymmetric aspects of the spectrum. In section \ref{Sec:string-D5} we review the status of the excitations of the string and the D5 brane configurations and fit the respective spectra into representations of $OSp(4^*|4)$. We conclude in section \ref{Sec:conclusions} with some open problems. We have relegated questions of conventions and more explicit calculations to a series of appendices.
\section{Review of the classical configurations}\label{Sec:Class-config}
In this section we briefly review the classical D3 brane configuration that describes the BPS Wilson loops we are interested in. It was first introduced in \cite{Drukker:2005kx}. Throughout the paper we will work in Euclidean signature but will tacitly switch to Lorentzian signature to discuss certain supersymmetry questions. We hope this will be clear to the reader from the context. See Appendix \ref{Appendix: Notation} for notation and conventions.
\subsection{$AdS_5\times S^5$ background}
The $AdS_5\times S^5$ type IIB background is described by a metric and a RR 5-form given by
\begin{empheq}{align}
    ds^2&=ds^2_{AdS_5}+L^2d\Omega_5^2,
    \\
    F_5&=-\frac{4}{L}\left(1+*\right)\textrm{vol}\left(AdS_5\right).
\end{empheq}
Both the $AdS_5$ space and the 5-sphere have radius $L$. Since the configurations we study in this paper are described by a D3 brane with $AdS_2\times S^2$ worldvolume, it is convenient to introduce coordinates that make this structure manifest. Following \cite{Yamaguchi:2006tq}, we consider a foliation of $AdS_5$ of the form
\begin{empheq}{align}\label{AdS5 Fibration}
    ds^2_{AdS_5}&=L^2\left(\cosh^2(u)ds^2_H+\sinh^2(u)d\Omega_2^2+du^2\right),
\end{empheq}
where $ds^2_H$ is the (unit) $AdS_2$ metric. It is clear that the induced geometry on the hypersurface $u=u_k$ corresponds to $AdS_2\times S^2$. In these coordinates the volume form reads
\begin{empheq}{align}
    \textrm{vol}\left(AdS_5\right)&=-L^5\cosh^2(u)\sinh^2(u)\,e^{\underline{0}}\wedge e^{\underline{1}}\wedge e^{\underline{2}}\wedge e^{\underline{3}}\wedge du,
\end{empheq}
where $(e^{\underline{0}},e^{\underline{1}})$ and $(e^{\underline{2}},e^{\underline{3}})$ are vielbeins for $AdS_2$ and $S^2$, respectively. Then the corresponding potential $C_4$ defined as
\begin{empheq}{align}
    F_5&=\left(1+*\right)dC_4
\end{empheq}
can be chosen to be
\begin{empheq}{align}
    C_4&=4L^4f(u)\,e^{\underline{0}}\wedge e^{\underline{1}}\wedge e^{\underline{2}}\wedge e^{\underline{3}}, \quad {\rm with} \quad  f(u)=\frac{1}{32}\sinh(4u)-\frac{u}{8}.
\end{empheq}
This potential is particularly useful since it lies entirely on $AdS_2\times S^2$. As explained in \cite{Drukker:2005kx}, different gauge choices for $C_4$ are related by a conformal transformation at the boundary of $AdS_5$.
\subsection{Classical D3 brane solutions}
The bosonic part of the D3 brane action in the is given by
\begin{empheq}{align}\label{Bosonic D3 Action}
    S_B&=T_{D3}\int d^4\sigma\,\sqrt{\textrm{det}\left(g+2\pi\alpha'F\right)}-T_{D3}\int P[C_4],
\end{empheq}
where $\sigma^{\alpha},\,\alpha=0,1,2,3$ are worldvolume coordinates, $P[\;]$ denotes the pullback to the worldvolume, $g_{\alpha\beta}=P[G]_{\alpha\beta}$ is the induced metric, and $F=dA$ is the field strength of the gauge field living on the brane. The tension of a D3 brane is
\begin{empheq}{align}
    T_{D3}&=\frac{N}{2\pi^2L^4}.
\end{empheq}

The classical configurations relevant to us are solutions to the equations of motion derived from \eqref{Bosonic D3 Action}. The boundary conditions are such that the hypersurface must pinch the boundary of $AdS_5$ along an infinite line or a circle, depending on the specific $\mathcal{N}=4$ SYM loop operator one is interested in. In either case, choosing a static gauge where the coordinates on $AdS_2$ and $S^2$ are identified with the worldvolume coordinates, the solution describing the BPS Wilson loop is given by\footnote{The electric field is imaginary in the Euclidean theory.}
\begin{empheq}{alignat=3}\label{Static Solution}
    u&=u_k,
    &\qquad \theta^{\hat{i}}&=\theta^{\hat{i}}_0,
    &\qquad  2\pi\alpha'F&=iL^2\cosh(u_k)e^{\underline{0}}\wedge e^{\underline{1}},
\end{empheq}
where $\theta^i$ are coordinates on $S^5$. The D-brane sits at a fixed point $u_k$ on the base space of the $AdS_5$ foliation. The value of this point is determined by the fundamental string charge $k$ dissolved on the brane by
\begin{empheq}{align}\label{u0value}
    \sinh(u_k)&=\frac{k\sqrt{\lambda}}{4N}\equiv\kappa.
\end{empheq}
The point $\theta^{\hat{i}}_0$ on the 5-sphere is arbitrary. The induced geometry is $AdS_2\times S^2$, with metric
\begin{empheq}{align}\label{Induced Geometry}
    ds^2&=L^2\left(\cosh^2(u_k)ds^2_H+\sinh^2(u_k)d\Omega_2^2\right).
\end{empheq}
Note that the radius of $AdS_2$ is $L\cosh(u_k)$ and the radius of $S^2$ is $L\sinh(u_k)$.

The difference between the solution dual to the infinite straight line and the solution dual to the circular Wilson loop lies in the global structure chosen to describe the $AdS_2$ space introduced in \eqref{AdS5 Fibration}. To see this, let the $AdS_2$ metric be described by the half-plane model,
\begin{empheq}{align}\label{AdS_2 Half-Plane}
    ds^2_H&=\frac{1}{r^2}\left(dx^2+dr^2\right).
\end{empheq}
Then the transformation $r^2=\rho^2+y^2$, $\sinh(u)=\rho/y$ brings the metric \eqref{AdS5 Fibration} to the form
\begin{empheq}{align}\label{AdS_5 Infinite Line}
    ds^2_{AdS_5}&=\frac{L^2}{y^2}\left(dx^2+d\rho^2+\rho^2d\Omega_2^2+dy^2\right).
\end{empheq}
We see that the embedding $y=\rho/\kappa$ pinches the boundary $y=0$ of $AdS_5$ along an infinite line spanned by the coordinate $x$. More explicitly, the induced metric takes the form
\begin{empheq}{align}\label{Indiced Metric Infinite Line}
    ds^2&=\frac{L^2\kappa^2}{\rho^2}\left(dx^2+(1+\frac{1}{\kappa^2})d\rho^2\right)+ L^2\kappa^2 d\Omega_2^2.
\end{empheq}
Thus we see that the holographic description of the infinite line Wilson loop is captured by the metric \eqref{AdS_2 Half-Plane}.

In contrast, the circular loop is better described by the disk model of $AdS_2$,
\begin{empheq}{align}\label{AdS_2 Disk}
    ds^2_H&=d\chi^2+\sinh^2\chi d\psi^2.
\end{empheq}
In this case the coordinate change $\cot\eta=\cosh(u)\sinh\chi$, $\coth\rho=\coth(u)\cosh\chi$ gives
\begin{empheq}{align}\label{AdS_5 Circle 1}
    ds^2_{AdS_5}&=\frac{L^2}{\sin^2\eta}\left(\cos^2\eta d\psi^2+d\rho^2+\sinh^2\rho d\Omega_2^2+d\eta^2\right).
\end{empheq}
This can be put in a more familiar form by writing $d\Omega_2^2=d\theta^2+\sin^2\theta d\phi^2$ and further defining
\begin{empheq}{align}\label{Transformation Circle 1->2}
    r_1&=\frac{R\cos\eta}{\cosh\rho-\sinh\rho\cos\theta}
    \\
    r_2&=\frac{R\sinh\rho\sin\theta}{\cosh\rho-\sinh\rho\cos\theta}
    \\
    y&=\frac{R\sin\eta}{\cosh\rho-\sinh\rho\cos\theta}
\end{empheq}
as was done in \cite{Drukker:2005kx}. Then,
\begin{empheq}{align}\label{AdS_5 Circle 2}
    ds^2_{AdS_5}&=\frac{L^2}{y^2}\left(dr_1^2+r_1^2d\psi+dr_2^2+r_2^2d\phi^2+dy^2\right).
\end{empheq}
The D3 brane worldvolume is now described by $\sin\eta=\sinh\rho/\kappa$. As we approach the boundary $y=0$ ($\eta=0$) the hypersurface becomes $r_2=0$, which corresponds to a circle of radius $r_1=R$ parameterized by $\psi$. Notice that the radius of the loop which appears explicitly in the transformation \eqref{Transformation Circle 1->2} does not appear in the metric \eqref{AdS_5 Circle 2} as a consequence of scale invariance.

One of the reasons we use the coordinate system \eqref{AdS5 Fibration} is that the infinite straight line and circular Wilson loops can be described in a unified way. Notice, however, that in the case of the circular loop, the solution only makes sense in Euclidean signature, since the metric \eqref{AdS_2 Disk} does not have a well defined Lorentzian counterpart. In contrast, the half-plane metric \eqref{AdS_2 Half-Plane} can be analytically continued to
\begin{empheq}{align}\label{AdS_2 Half-Plane Lorentzian}
    ds^2_H&=\frac{1}{r^2}\left(-dt^2+dr^2\right).
\end{empheq}
In this case the gauge field becomes
\begin{empheq}{alignat=3}\label{Gauge Field Lorentzian}
    2\pi\alpha'F&=\frac{L^2\cosh(u_k)}{r^2}dt\wedge dr.
\end{empheq}

The solutions possess a $SL(2,\mathds{R})\times SO(3)\times SO(5)$ symmetry corresponding to isometries of the worldvolume geometry and rotations of $S^5$ about a fixed point. These coincide with the bosonic symmetries preserved by the Wilson loops in the gauge theory dual. It is also shown in \cite{Drukker:2005kx} that these D3 branes preserve half of the target space supersymmetries. We will come back to this in section \ref{Supersymmetry}.
\section{Open string excitations}\label{Sec:D3}
In this section we consider the open string fluctuations of the corresponding Born-Infeld (BI) action. This is, by now, a rather mature subject in the context of the AdS/CFT correspondence. A unifying theme of this study is the fact that some probe branes have worldvolumes containing an $AdS_p$ factor, pointing to the possibility of some effective conformal theory different from the original ${\cal N}=4$ SYM. Let us briefly review the evolution of the subject to see where our example fits. One of the first examples was provided by \cite{DeWolfe:2001pq} in the context of a defect CFT, recall that the worldvolume of the probe D5 brane in that case is $AdS_4\times S^2$. Perhaps a more widely known example of this class is given by the study of a probe $D7$ brane whose worldvolume is $AdS_5\times S^3$ \cite{Kruczenski:2003be} where the application to ${\cal N}=2$ SYM with fundamental matter was highlighted.

There have been many examples of the above situation. A cleaner conceptual framework, however, arose clearly in the works \cite{Karch:2005ms,Arean:2006pk}. These works provided a holographic renormalization description of worldvolume probe branes yielding a recipe for how to read the dimensions of operators dual to modes coming from the defect brane or probe brane. Interestingly, the fields in this brane are not the ${\cal N}=4$ fields which live in the original D3.  Due to the crucial role of asymptotic data, it turns out that solutions to the DBI action are essentially classified as solution of free fields in $AdS_p\times S^q$ which is the worldvolume of the probe branes. According to \cite{Karch:2005ms} the open string modes emerging from the BI action are effectively described by a scalar in $AdS_{d+1}$ with action
\begin{empheq}{align}
    S&=\frac{1}{2}\int d^{d+1}x\sqrt{g}\left(g^{\alpha\beta}\partial_{\alpha}\Phi\partial_{\beta}\Phi+M^2\Phi^2\right).
\end{empheq}
This scalar is dual to some gauge-invariant CFT operator with dimension $\Delta$ given by $M^2 =\Delta(\Delta-d)$. More importantly for us, the corresponding operator is an operator in the defect theory. A key point in  \cite{Karch:2005ms} is that in the cases they considered (embedding without worldvolume fluxes) the counterterms for the DBI action are identical to the counterterms for a free scalar in $AdS$.

Our work provides a further generalization where the probe brane has flux in its worldvolume. We do not work out the general case similar to \cite{Karch:2005ms}, we defer this analysis to the future. It suffices to say that we also obtain that our open string fluctuations are described by fields in $AdS_p$ albeit with a metric different from the induced metric. Let us elaborate on this point.

The D3 brane fluctuations around the static solution \eqref{Static Solution} are described by a field theory living on the worldvolume of the brane. In the absence of a background flux $F_{\alpha\beta}$ the natural geometry is given by the induced metric, i.e.,  the pullback of $AdS_5\times S^5$ to the worldvolume of the brane. As we have seen in (\ref{Induced Geometry}), this is $AdS_2\times S^2$ with radii $L\cosh(u_k)$ and $L\sinh(u_k)$. As explained in \cite{Martucci:2005rb}, one of the effects of adding a flux is to deform the geometry according to
\begin{empheq}{align}\label{Definition of Deformed Metric}
    \hat{g}_{\alpha\beta}&=g_{\alpha\beta}-F_{\alpha\gamma}g^{\gamma\delta}F_{\delta\beta}
\end{empheq}
This deformation is crucial in casting the fermionic part of the action in a canonical form. For the case at hand we readily find
\begin{empheq}{align}\label{Deformed Geometry}
    d\hat{s}^2&=L^2\sinh^2(u_k)\left(ds^2_H+d\Omega_2^2\right)
\end{empheq}
so that the metric is still given by $AdS_2\times S^2$ but with equal radii $L\sinh(u_k)$. In this sense the effect of the worldvolume flux is rather innocuous. Of course, as we will discuss below, there are other less trivial consequences. In particular we can infer that changing the radius of $AdS_2$, from the holographical point of view amounts to changing the conformal dimension of the dual operators.
\subsection{The D3 brane action in $AdS_5\times S^5$}
The bosonic part of the D3 brane action in the $AdS_5\times S^5$ background is given equation \eqref{Bosonic D3 Action}, which we reproduce here for convenience,
\begin{empheq}{align}\label{Bosonic D3 Action Again}
    S_B&=T_{D3}\int d^4\sigma\,\sqrt{\textrm{det}\left(g+2\pi\alpha'F\right)}-T_{D3}\int P[C_4].
\end{empheq}
The construction of a quadratic fermionic action was presented in a series of interesting works by Martucci \cite{Marolf:2003ye,Marolf:2003vf,Martucci:2005rb,Martucci:2005ht,Martucci:2006ij}. Here we will closely follow  the notation and presentation of \cite{Martucci:2005rb}. For the $AdS_5\times S^5$ background the action reduces to
\begin{empheq}{align}\label{Fermionic D3 Action}
    S_F&=\frac{T_{D3}}{2}\int d^4\sigma\,\sqrt{\textrm{det}\left(g+2\pi\alpha'F\right)}\,\overline{\Theta}\left(1-\Gamma_{D3}\right)\tilde{M}^{\alpha\beta}\Gamma_{\beta}D_{\alpha}\Theta.
\end{empheq}
Here $\Theta$ is a doublet of 10d positive chirality Majorana-Weyl spinors, $\Gamma_{\alpha}=\partial_{\alpha}x^m\Gamma_m$ is the pullback of the spacetime Dirac matrices, $\tilde{M}^{\alpha\beta}$ is the inverse of
\begin{empheq}{alignat=2}\label{Definition of tilde M}
    \tilde{M}_{\alpha\beta}&=g_{\alpha\beta}+2\pi\alpha'F_{\alpha\beta}\tilde{\Gamma}
    &\qquad
    \tilde{\Gamma}&=\Gamma^{11}\otimes\sigma_3
\end{empheq}
and $\Gamma_{D3}$ is a projector ensuring invariance of the action under $\kappa$-symmetry (see equation \eqref{Definition of Gamma_D3}). Also, $D_{\alpha}=\partial_{\alpha}x^mD_m$ is the pullback of the type IIB covariant derivative, which in our case reads
\begin{empheq}{align}\label{IIB Covariant Derivative 10d}
    D_m&=\nabla_m+\frac{1}{16}\slashed{F}_{(5)}\Gamma_m\otimes\left(i\sigma_2\right).
\end{empheq}

Technically the action in \cite{Martucci:2005rb} is defined in Lorentzian signature. In particular, the Majorana condition changes under Euclidean continuation. For our purposes, we can still think of fermions being defined in Lorentzian signature and simply replace $t=ix$ when appropriate.

As shown in \cite{Martucci:2005rb}, the complete D-brane action is invariant under (linearized) supersymmetry transformations induced by the existence of target space Killing spinors. Since the classical embeddings considered here preserve half of the $AdS_5\times S^5$ supersymmetries, we expect the action for the quadratic fluctuations around these backgrounds to be supersymmetric. Instead of verifying this explicitly, we will show in the next section that the spectrum of excitations falls into multiplets of the appropriate supergroup.
\subsection{Bosonic fluctuations}
The local symmetries of the complete D-brane action include worldvolume diffeomorphisms and $\kappa$-symmetry \cite{Martucci:2005rb}. Let us comment on the gauge fixing procedure of diffeomorphisms following \cite{Drukker:2000ep}, \cite{Martucci:2005rb}. We will discuss $\kappa$-fixing in the next section.

Suppose we have a particular embedding $x^m(\sigma)$ that solves the Dp-brane equations of motion. The standard (static) gauge condition consists of fixing $x^{\alpha}(\sigma)=\sigma^{\alpha}$ for $p$ of the spacetime coordinates. Then, when considering fluctuations $\delta x^m(\sigma)$ around the solution, we should impose $\delta x^{\alpha}=0$ and consider the transverse fluctuations $\delta x^{\hat{m}}$ as physical. We will adopt this static gauge in what follows.

We would like, however, to briefly comment on a more geometrical gauge fixing procedure outlined in \cite{Martucci:2005rb}. Consider a target space vielbein $E^{\underline{m}}=(E^{\underline{\alpha}}, E^{\underline{\hat{m}}})$, such that the pull-back of $E^{\underline{\alpha}}$ to the worldvolume form a vielbein for the induced geometry while the pulled-back $E^{\underline{\hat{m}}}$ vanish. This explicitly breaks the local Lorentz invariance of the theory to $SO(p+1)\times SO(9-p)$. We can then consider the tangent space fluctuations
\begin{equation}
    \phi^{\underline{\hat{m}}}=E^{\underline{\hat{m}}}_{\phantom{\underline{\hat{m}}}m}\delta x^m
\end{equation}
as our worldvolume fields and fix the diffeomorphism invariance by the condition
\begin{equation}
    E^{\underline{\alpha}}_{\phantom{\underline{\alpha}}m}\delta x^m=0.
\end{equation}
The choice of $AdS_5$ coordinates in \eqref{AdS5 Fibration} actually makes this method equivalent to choosing static gauge. However, this gauge fixing condition is better suited to be used in more general coordinates, such as global or Poincare coordinates in $AdS$.

Following the above discussion, we choose a gauge such that the coordinates along $AdS_2\times S^2$ do not fluctuate and expand the remaining bosonic fields as
\begin{empheq}{alignat=3}
    u&=u_k+\phi^{4}
    &\qquad
    \theta^{\hat{i}}&=\theta^{\hat{i}}_0+\phi^{\hat{i}}
    &\qquad
    2\pi\alpha'A&=2\pi\alpha' A_0+a.
\end{empheq}
Notice that we have absorbed a factor of $2\pi\alpha'$ in the gauge field $a$. The corresponding tangent space fluctuations are
\begin{empheq}{alignat=2}
    \phi^{\underline{4}}&=L\phi^{4},
    &\qquad
    \phi^{\underline{\hat{i}}}&=Le^{\underline{\hat{i}}}_{\phantom{\underline{\hat{i}}}\hat{i}}\phi^{\hat{i}}.
\end{empheq}

The quadratic action for the perturbations is obtained by expanding the bosonic action \eqref{Bosonic D3 Action Again} to second order. We present the details of this calculation in Appendix \ref{Appendix: Explicit Calculations}. Our result is
\begin{empheq}{align}
    S_B&=S_B^{(0)}+S_B^{(1)}+S_B^{(2)}+\cdots,
\end{empheq}
where $S^{(0)}$ is the on-shell action for the background solution and
\begin{empheq}{align}\label{Quadratic Bosonic Action}
    S_B^{(2)}&=S_{\phi}^{(2)}+S_a^{(2)},
    \\\nonumber
    S_{\phi}^{(2)}&=\frac{T_{D3}\coth(u_k)}{2}\int d^4\sigma\sqrt{\hat{g}}\hat{g}^{\alpha\beta}\left(\partial_{\alpha}\phi^{\underline{\hat{4}}}\partial_{\beta}\phi_{\underline{\hat{4}}}+\partial_{\alpha}\phi^{\underline{\hat{i}}}\partial_{\beta}\phi_{\underline{\hat{i}}}\right),
    \\\nonumber
    S_a^{(2)}&=\frac{T_{D3}\coth(u_k)}{4}\int d^4\sigma\sqrt{\hat{g}}\hat{g}^{\alpha\beta}\hat{g}^{\gamma\delta}f_{\alpha\gamma}f_{\beta\delta}.
\end{empheq}
where $f_{\alpha\beta}=\partial_\alpha a_\beta -\partial_\beta a_\alpha$. As expected, the linear terms $S_B^{(1)}$ vanish by virtue of the equations of motion. In the above expressions we obtain, naturally, the deformed $AdS_2\times S^2$ metric \eqref{Deformed Geometry}.

According to the AdS/CFT dictionary  $\exp(-S_B^{(0)})$ gives the expectation value of the Wilson loop in the t' Hooft  limit \cite{Maldacena:1998im,Rey:1998ik}. The on-shell action $S_B^{(0)}$ is divergent and must be renormalized to render a finite result. A systematic prescription was proposed in \cite{Drukker:1999zq} and applied to the D3 brane configuration in \cite{Drukker:2005kx}.

The fact that the fluctuation $\phi^{\underline{4}}$ is massless is a consequence of supersymmetry; the worldvolume flux ensures that the mass term coming from the DBI part of the action exactly cancels the contribution of the WZ term. This cancelation is shown explicitly in appendix \ref{Appendix: Explicit Calculations}.
\subsection{Fermionic fluctuations}
In the context of AdS/CFT fluctuations of fermions have played a relatively secondary role. To our knowledge, there is only one explicit computation of the fermionic spectrum of a D7-brane in $AdS_5\times S^5$ \cite{Kirsch:2006he}, but without worldvolume fluxes.  Some of the classical brane configurations that appear in this work have cousins in the context of confining theories where they describe confining $k$-strings which are bound states of $k$ quarks and $k$ anti-quarks. The study of fluctuations in that context is important for the computation of the L\"uscher term. The works \cite{PandoZayas:2008hw,Doran:2009pp,Stiffler:2009ma,Stiffler:2010pz} presented a study of the fluctuations and found certain universality in the value of the L\"uscher term.

Since the fermionic fields vanish in the classical solution \eqref{Static Solution}, we can consider $\Theta$ in \eqref{Fermionic D3 Action} as the fermionic fluctuation. Moreover, all the bosonic quantities can be evaluated on the background. The details of the calculation of the fermionic action are shown in Appendix \ref{Appendix: Explicit Calculations}. We find that
\begin{empheq}{align}
    S_{\Theta}^{(2)}=\frac{T_{D3}\coth(u_k)}{2}\int d^4\sigma\sqrt{\hat{g}}\,\overline{\Theta}\left(1+\Gamma_{D3}^{(0)}\right)\hat{\slashed{\nabla}}\Theta.
\end{empheq}
In this expression,
\begin{empheq}{align}
    \Gamma_{D3}^{(0)}&=\Gamma_{\underline{0123}}\otimes\left(i\sigma_2\right)
\end{empheq}
is the $\kappa$-symmetry projector and $\hat{\nabla}_{\alpha}$ is the covariant derivative with respect to the deformed geometry \eqref{Deformed Geometry}. More explicitly,
\begin{empheq}{align}
    \hat{\nabla}_{\alpha}&=\partial_{\alpha}+\frac{1}{4}\hat{w}^{\underline{\beta\gamma}}_{\phantom{\underline{\alpha\beta}}\alpha}\Gamma_{\underline{\beta\gamma}}
\end{empheq}
where $\hat{w}^{\underline{\alpha\beta}}$ denotes the spin connection of the deformed metric $\hat{g}_{\alpha\beta}$ and $\Gamma_{\underline{\alpha}}$ are 10d gamma matrices. The Dirac operator $\hat{\slashed{\nabla}}=\hat{\Gamma}^{\alpha}\hat{\nabla}_{\alpha}$ is defined using $\hat{\Gamma}_{\alpha}=\hat{e}^{\underline{\alpha}}_{\phantom{\underline{\alpha}}\alpha}\Gamma_{\underline{\alpha}}$ and $\hat{\Gamma}^{\alpha}=\hat{g}^{\alpha\beta}\hat{\Gamma}_{\beta}$, with $\hat{e}^{\underline{\alpha}}$ being a vielbein for $\hat{g}_{\alpha\beta}$.

To fix the local $\kappa$-symmetry we use the prescription of \cite{Martucci:2005rb}, which is,
\begin{empheq}{alignat=2}
    \tilde{\Gamma}\Theta&=\Theta,
    &\qquad
    \tilde{\Gamma}&=\Gamma^{11}\otimes\sigma_3
\end{empheq}
This sets the lower component of $\Theta$ to zero. Denoting the upper component also as $\Theta$, the gauge fixed action reads
\begin{empheq}{align}
    S_{\Theta}^{(2)}=\frac{T_{D3}\coth(u_k)}{2}\int d^4\sigma\sqrt{\hat{g}}\,\overline{\Theta}\hat{\slashed{\nabla}}\Theta.
\end{empheq}

Now, under $SO(9,1)\rightarrow SO(3,1)\times SO(6)$ the Majorana-Weyl spinor $\Theta$ is decomposed into a $SO(3,1)$ Weyl spinor $\Theta_A$ transforming as a $\mathbf{4}$ of $SO(6)\simeq SU(4)$. Further decomposing $SO(6)\rightarrow SO(5)$, to accommodate for the symmetries of the Wilson loop, we get a $\mathbf{4}$ of $SO(5)\simeq USp(4)$. This way we obtain the fermionic action
\begin{empheq}{align}\label{Quadratic Fermionic Action}
    S_{\Theta}=T_{D3}\coth(u_k)\int d^4\sigma\sqrt{\hat{g}}\,\overline{\Theta}_A\hat{\slashed{\nabla}}\Theta_A,
\end{empheq}
where the covariant derivative is the same as before but with 4-dimensional Dirac matrices.

Naively, we would have expected a mass term coming from the coupling to $\slashed{F}_5$ in \eqref{IIB Covariant Derivative 10d}. However, supersymmetry conspires to precisely cancel this term against the contributions coming from the extrinsic curvature in the 10d spin connection (see Appendix \ref{Appendix: Explicit Calculations} for details).
\subsection{Compactification on $S^2$}\label{Subsection: Compactification}
Since the D3 brane worldvolume has the product structure $AdS_2\times S^2$ we can compactify on the sphere. We now display the effective $AdS_2$ theory in agreement with general expectations of the string theory description. The resulting 2d actions are given by (see Appendix \ref{Appendix: Compactification} for all the relevant explicit details)
\begin{empheq}{align}
    S_{\phi}^{(2)}&=\frac{T_{D3}}{2}\coth(u_k)\sum_{l=0}^{\infty}\sum_{m=-l}^l\int d^2\sigma\sqrt{\hat{g}}\left(\hat{g}^{\mu\nu}\partial_{\mu}\phi^{\underline{\hat{4}}}_{lm}\partial_{\nu}\phi^{\underline{\hat{4}}}_{lm}+\frac{l(l+1)}{L^2\sinh^2(u_k)}\phi^{\underline{\hat{4}}}_{lm}\phi^{\underline{\hat{4}}}_{lm}\right)
    \\\nonumber
    &+\frac{T_{D3}}{2}\coth(u_k)\sum_{l=0}^{\infty}\sum_{m=-l}^l\int d^2\sigma\sqrt{\hat{g}}\left(\hat{g}^{\mu\nu}\partial_{\mu}\phi^{\underline{\hat{i}}}_{lm}\partial_{\nu}\phi^{\underline{\hat{i}}}_{lm}+\frac{l(l+1)}{L^2\sinh^2(u_k)}\phi^{\underline{\hat{i}}}_{lm}\phi^{\underline{\hat{i}}}_{lm}\right),
    \\
    S_a^{(2)}&=\frac{T_{D3}}{4}\coth(u_k)\sum_{l=0}^{\infty}\sum_{m=-l}^l\int d^2\sigma\sqrt{\hat{g}}\left(\hat{g}^{\mu\nu}\hat{g}^{\rho\sigma}f^{lm}_{\mu\rho}f^{lm}_{\nu\sigma}+\frac{2l(l+1)}{L^2\sinh^2(u_k)}g^{\mu\nu}a^{lm}_{\mu}a^{lm}_{\nu}\right)
    \\\nonumber
    &+\frac{T_{D3}}{2}\coth(u_k)\sum_{l=1}^{\infty}\sum_{m=-l}^l\int d^2\sigma\sqrt{\hat{g}}\left(\hat{g}^{\mu\nu}\partial_{\mu}a_{lm}\partial_{\nu}a_{lm}+\frac{l(l+1)}{L^2\sinh^2(u_k)}\left(a_{lm}\right)^2\right),
    \\
    S_{\Theta}^{(2)}&=2T_{D3}\coth(u_k)\sum_{l=\frac{1}{2}}^{\infty}\sum_{m=-l}^{l}\int d^2\sigma\sqrt{\hat{g}}\bar{\Theta}_A^{lm}\left(\hat{\slashed{\nabla}}+\frac{i\left(l+\frac{1}{2}\right)\gamma}{L\sinh(u_k)}\right)\Theta_A^{lm},
\end{empheq}
where $f_{\mu\nu}^{lm}=\partial_{\mu}a^{lm}_{\nu}-\partial_{\nu}a^{lm}_{\mu}$. All the geometric quantities appearing above are intrinsically 2-dimensional and are defined in terms of the $AdS_2$ factor of the deformed metric \eqref{Deformed Geometry}. In particular, the Dirac matrices $\gamma_{\underline{\mu}}$ implicit in the fermionic action are $2\times2$ matrices. Also, $\gamma=\gamma_{\underline{01}}$.

These expressions follow from the expansion of the 4-dimensional fields in terms of scalar, vector and spinor harmonics on $S^2$. In the case of $\theta^{lm}_A$, which are 2d Dirac spinors, the quantum number $l$ takes values $l=\frac{1}{2},\frac{3}{2},\ldots$, as appropriate for fermions. Notice that the scalar modes $a_{lm}$ coming from the gauge field components along the sphere start at $l=1$. Also, the $l=0$ mode of the gauge field is massless so it has no propagating degrees of freedom.
\section{Supersymmetry}\label{Supersymmetry}
In this section we discuss the symmetries of the BPS Wilson loops and how the spectrum of open string fluctuations fits into representations of the corresponding super group. Since we will be discussing supersymmetry, we switch to Lorentzian signature.
\subsection{Symmetries of the Wilson loop}
The $\mathcal{N}=4$ SYM theory has a supersymmetry group given by $SU(2,2|4)$. The bosonic symmetries are $SU(2,2)\times SU(4)$, where $SU(2,2)\simeq SO(4,2)$ is the conformal group in four dimensions and the $SU(4)\simeq SO(6)$ factor acts as an $R$-symmetry. In the string theory description, these symmetries are realized as isometries of $AdS_5\times S^5$.

Let us review the subgroup of $SU(2,2|4)$ preserved by the straight line Wilson loop. This is done in detail in \cite{Gomis:2006sb}. First we recall that a general bosonic Wilson loop operator in a representation $R$ of $SU(N)$ is defined as
\begin{empheq}{align}
    W_R(C)&=\textrm{Tr}_RP\exp\left(i\int_Cds\left(A_{\mu}\dot{x}^{\mu}+\phi_I\dot{y}^I\right)\right),
\end{empheq}
where $C$ labels a curve $\left(x^{\mu}(s),y^I(s)\right)$ in $\mathcal{N}=4$ superspace, $I$ is a vector index of $SO(6)$, and $P$ denotes path ordering along the loop. As shown in \cite{Gomis:2006sb}, in order to preserve supersymmetry, the curve $x^{\mu}(s)$ must be an infinite timelike line, which we parameterize by $x^{\mu}(s)=\left(x^0(s),0,0,0\right)$. Supersymmetry also implies that $\dot{y}^I=n^I$, where $n^I$ is a constant unit vector in $\mathds{R}^6$.

Now, acting on the spacetime coordinates, the generators $\left(P_{\mu},J_{\mu\nu},D,K_{\mu}\right)$ of $SO(4,2)$ read
\begin{empheq}{align}
    \delta x^{\mu}&=a^{\mu}+w^{\mu}_{\phantom{\mu}\nu}x^{\nu}+\lambda x^{\mu}+b^{\mu}x^2-2b\cdot xx^{\mu}
\end{empheq}
where $\left(a^{\mu},w^{\mu}_{\phantom{\mu}\nu},\lambda, b^{\mu}\right)$ are the corresponding transformation parameters. Conserving the form of the loop imposes the conditions
\begin{empheq}{alignat=3}
    a^i&=w^i_{\phantom{i}0}&=b^i&=0
\end{empheq}
Thus, the subgroup preserved by the Wilson loop is generated by $\left(P_0,J_{ij},D,K_0\right)$. The interpretation of these transformations is simple: an infinite line is left invariant by translations along the line, rotations around the line, and dilatations of the coordinates. The operator $K_0$ generates a special conformal transformation. The generators $J_{ij}$ span the $SU(2)\simeq SO(3)$ algebra while the rest satisfy
\begin{empheq}{alignat=3}
    [P_0,K_0]&=-2D,
    &\quad
    [P_0,D]&=-P_0,
    &\quad
    [K_0,D]&=K_0.
\end{empheq}
This is the $SU(1,1)\simeq SO(2,1)\simeq SL(2,\mathds{R})$ algebra. Thus, we see that the infinite line preserves a $SO(4^*)\simeq SL(2,\mathds{R})\times SO(3)$ subgroup of $SO(4,2)$. Finally, the choice of a vector $n^I$ breaks the $SO(6)$ $R$-symmetry down to $SO(5)\simeq USp(4)$.

The infinite line Wilson loop also preserves 16 of the 32 supersymmetries of $SU(2,2|4)$. The original works of \cite{Nahm:1977tg,Gunaydin:1986fe} identified all the possible $AdS$ supegroups and their multiplets. Among the subgroups of $SU(2,2|4)$, the supergroup $OSp\left(4^*|4\right)$ has $SL(2,\mathds{R})\times SO(3)\times SO(5)$ as its even subgroup and 16 fermionic generators.

Turning to the holographic description of the BPS Wilson loops, we see that the classical D3 brane solutions \eqref{Static Solution} possess a $SL(2,\mathds{R})\times SO(3)\times SO(5)$ symmetry corresponding to isometries of the worldvolume geometry and rotations of $S^5$ about a fixed point. It is also shown in \cite{Drukker:2005kx} that these D3 brane configurations preserve half of the target space supersymmetries. As expected, these coincide with the symmetries preserved by the Wilson loop in the gauge theory dual.
\subsection{Conformal Dimensions}
As we have seen in section \ref{Subsection: Compactification}, the bosonic open string fluctuations are described by scalar and vector fields in $AdS_2$, all with masses
\begin{empheq}{alignat=2}
    m_l^2&=l(l+1)/L^2\sinh^2(u_k)
    &\qquad
    l&=0^*,1,\ldots
\end{empheq}
where $^*$ reminds us that some fluctuations do not include the $l=0$ mode. According to the standard AdS/CFT dictionary, the conformal dimensions of the operators dual to such modes are given by the formulas
\begin{empheq}{alignat=2}
    h^{scalar}_{\pm}&=\frac{1}{2}\left(d\pm\sqrt{d^2+4m^2R^2}\right)
    &\qquad
    h^{vector}_{\pm}&=\frac{1}{2}\left(d\pm\sqrt{(d-2)^2+4m^2R^2}\right)
\end{empheq}
where $R$ is the radius of $AdS_{d+1}$. In our case $d=1$ and $R=L\sinh(u_k)$, so
\begin{empheq}{alignat=2}
    h&=l+1
    &\qquad
    l&=0^*,1,\ldots
\end{empheq}
for all the bosonic fields. Similarly, from the formula
\begin{empheq}{align}
    h^{spinor}&=\frac{d}{2}+|m|R
\end{empheq}
we see that the fermionic modes $\Theta_A^{lm}$ have
\begin{empheq}{alignat=2}
    h&=l+1
    &\qquad
    l&=\frac{1}{2},\frac{3}{2},\ldots
\end{empheq}

Notice that the masses depend on the radius of $S^2$ while the formulas for the conformal dimensions involve the $AdS_2$ radius. The fact that the perturbations see the deformed metric \eqref{Deformed Geometry} instead of the induced metric \eqref{Induced Geometry} is crucial to get rational values for $h$. This is a consequence of the supersymmetry preserved by the D3 brane, as we will see below.

All in all, the spectrum of excitations of the D3 brane is given by a KK tower of fields propagating in $AdS_2$ labeled by their $SL(2,\mathds{R})\times SO(3)\times SO(5)$ quantum numbers. This result is summarized in table \ref{Table: Full D3 Spectrum}. At the lowest level there are six massless and six massive (two triplets of $SO(3)$) bosonic modes. This spectrum is quite different from the expectations based on the calculation using fundamental strings, where the counting was five massless and three massive modes \cite{Forste1999,Drukker:2000ep,Sakaguchi:2007ea,Kruczenski:2008zk}.
\begin{table}[h]
  \centering
  \begin{tabular}{|c|c|c|c|c|c|c|}
    \hline
    \multicolumn{2}{|c|}{2d field} & 4d origin & $SL(2,\mathds{R})$ & $SO(3)$ & $SO(5)$ & \\
    \hline\hline
    \multirow{4}{*}{Bosons} & $\phi^{\underline{4}}_{lm}$ & embedding in $AdS_5$ & $l+1$ & $l$ & $\mathbf{1}$ & $l\geq0$ \\
    \cline{2-7}
     & $\phi^{\underline{\hat{i}}}_{lm}$ & embedding in $S^5$ & $l+1$ & $l$ & $\mathbf{5}$ & $l\geq0$ \\
    \cline{2-7}
     & $a_{\mu}^{lm}$ & gauge field along $AdS_2$ & $l+1$ & $l$ & $\mathbf{1}$ & $l\geq1$ \\
    \cline{2-7}
     & $a_{lm}$ & gauge field along $S^2$ & $l+1$ & $l$ & $\mathbf{1}$ & $l\geq1$ \\
    \hline\hline
    Fermions & $\Theta_A^{lm}$ & IIB spinor & $l+1$ & $l$ & $\mathbf{4}$ & $l\geq\frac{1}{2}$ \\
    \hline
  \end{tabular}
  \caption{{\small KK tower of modes and their transformation properties under $SL(2,\mathds{R})\times SO(3)\times SO(5)$. The representations of $SL(2,\mathds{R})$ are labeled by the $L_0=h$ eigenvalue of the highest weight state.}}\label{Table: Full D3 Spectrum}
\end{table}
\subsection{Supersymmetry of the spectrum}
We are interested in understanding how the excitations we have described can be organized in representations of supersymmetry. Similar fittings of KK modes into $AdS$ supermultiplets have appeared in the context of the AdS/CFT correspondence. In particular, compactifications of supergravity theories on $AdS_2\times S^2$ have been presented thoroughly in \cite{deBoer:1998ip,Michelson:1999kn,Corley:1999uz,Lee:1999yu}. In these examples the relevant supergroup is $SU(1,1|2)$, which has $SL(2,\mathds{R})\times SO(3)$ as its even subgroup. In our case we have an extra $SO(5)$ symmetry which makes $OSp(4^*|4)$ the relevant supergroup. The spectrum of open string fluctuations should then fall into multiplets of $OSp(4^*|4)$.

Lowest weight representations of the super-group $OSp(2m^*|2n)$ where studied in \cite{Gunaydin1991}. In the case of $OSp(4^*|4)$, the so-called doubleton representations can be labeled by a half-integer $j$ and have the following $SO(4^*)\times USp(4)\simeq SL(2,\mathds{R})\times SU(2)\times SO(5)$ content (see Appendix \ref{Appendix: OSp(4*|4) Algebra}):
\begin{empheq}{align}\label{OSp D3 multiplet}
    \mathbf{j}&=(j+1,j,\mathbf{5})\oplus(j+{\textstyle\frac{3}{2}},j+{\textstyle\frac{1}{2}},\mathbf{4})\oplus(j+2,j+1,\mathbf{1})
    \\\nonumber
    &\oplus(j+{\textstyle\frac{1}{2}},j-{\textstyle\frac{1}{2}},\mathbf{4})\oplus(j+1,j,\mathbf{1})
    \\\nonumber
    &\oplus(j,j-1,\mathbf{1}),
\end{empheq}
for $j\geq1$ and
\begin{empheq}{align}
    \label{OSp D3 multiplet j=0}
    \mathbf{0}&=(1,0,\mathbf{5})\oplus({\textstyle\frac{3}{2}},{\textstyle\frac{1}{2}},\mathbf{4})\oplus(2,1,\mathbf{1}),
    \\
    \label{OSp D3 multiplet j=1/2}
    {\textstyle\mathbf{\frac{1}{2}}}&=({\textstyle\frac{3}{2}},{\textstyle\frac{1}{2}},\mathbf{5})\oplus(2,1,\mathbf{4})\oplus({\textstyle\frac{5}{2}},{\textstyle\frac{3}{2}},\mathbf{1})
    \\\nonumber
    &\oplus(1,0,\mathbf{4})\oplus({\textstyle\frac{3}{2}},{\textstyle\frac{1}{2}},\mathbf{1}).
\end{empheq}
for the smallest multiplets.

The guiding principle in identifying the different states in the multiplet is their $SO(5)$ representation. Looking at table \ref{Table: Full D3 Spectrum}, we first notice that only multiplets with integer $j$ can occur in the spectrum. It is also clear that the scalar excitations in $S^5$ correspond to the first state in \eqref{OSp D3 multiplet}; these are the only fields that transform as a $\mathbf{5}$ of $SO(5)$. By looking at the first multiplet \eqref{OSp D3 multiplet j=0} we see that the third state must correspond to a bosonic fluctuation whose excitations start at $j=1$, i.e. one of the two gauge field fluctuations. Looking at the next multiplet,
\begin{empheq}{align}\label{OSp D3 multiplet j=1}
    \mathbf{1}&=\left(2,1,\mathbf{5}\right)\oplus\left(5/2,3/2,\mathbf{4}\right)\oplus\left(3,2,\mathbf{1}\right)
    \\\nonumber
    &\oplus\left(3/2,1/2,\mathbf{4}\right)\oplus\left(2,1,\mathbf{1}\right)
    \\\nonumber
    &\oplus\left(1,0,\mathbf{1}\right)
\end{empheq}
we realize that the fifth $\left(2,1,\mathbf{1}\right)$ and sixth $\left(1,0,\mathbf{1}\right)$ states must be identified with the other gauge field fluctuation and the scalar excitation in $AdS_5$, respectively. In fact, this identification works for any integer $j$. In table \ref{Table: Full D3 Spectrum}, this simply amounts to relabeling $l=j-1$ for $\phi^{\underline{4}}_{lm}$, $l=j$ for $\phi^{\underline{\hat{i}}}_{lm}$, $l=j$ for $a_{\mu}^{lm}$, and $l=j+1$ for $a_{lm}$. This implies that the lowest lying modes do not fit in a single multiplet, rather they are split among an entire $\mathbf{0}$ and part of a $\mathbf{1}$ multiplet.

Now, recall that the fermionic fluctuations $\Theta_A^{lm}$ in table \ref{Table: Full D3 Spectrum} are Dirac spinors. By decomposing them into two real spinors and identifying $l=j+\frac{1}{2}$ for one and $l=j-\frac{1}{2}$ for the other, we get the same fermionic content as \eqref{OSp D3 multiplet}.

In summary, we find that the $OSp(4^*|4)$ structure of the spectrum of excitations is
\begin{empheq}{align}
    \bigoplus_{j\geq0}\mathbf{j}
\end{empheq}
where the multiplets $\mathbf{j}$ are given by \eqref{OSp D3 multiplet}, \eqref{OSp D3 multiplet j=0} and \eqref{OSp D3 multiplet j=1/2}.
\section{Holographic Excitations: Fundamental String and D5 Brane}\label{Sec:string-D5}
Having discussed the case of the probe D3 brane in detail, we would like to complete the analysis of excitations of holographic Wilson loops by discussing the fundamental string and the D5 brane solutions. As explained in \cite{Gomis:2006im,Gomis:2006sb,Yamaguchi:2006tq}, the D3 brane describes the Wilson loop in the symmetric representation of $SU(N)$ while the D5 corresponds to the antisymmetric representation. The description in terms of a string captures the fundamental representation of $SU(N)$.
\begin{table}[h]
  \centering
  \begin{tabular}{|c|c|l|l|c|}
    \hline
    Configuration & Representation & Worldvolume & Isometries & Supergroup \\
    \hline\hline
    F1 & Fundamental & $AdS_2$ & $SL(2,\mathds{R})$ & \\
    \cline{1-4}
    D3 & Symmetric & $AdS_2\times S^2$ & $SL(2,\mathds{R})\times SO(3)$ & {\large$OSp(4^*|4)$} \\
    \cline{1-4}
    D5 & Antisymmetric & $AdS_2\times S^4$ & $SL(2,\mathds{R})\times SO(5)$ & \\
    \hline
  \end{tabular}
  \caption{BPS Wilson loops in various representations and their holographic descriptions.}\label{}
\end{table}

From the field theory perspective, the symmetries of the Wilson loop operator do not depend on the particular representation of the gauge group. We therefore expect that the excitations of the fundamental string and D5 brane dual to the Wilson loop operators also fall into representations of the supergroup $OSp(4^*|4)$, as in the case of the D3 brane.
\subsection{Fundamental String}
The systematic study of the semi-classical fluctuations of strings dual to the BPS Wilson loops was carried out in \cite{Forste1999,Drukker:2000ep,Sakaguchi:2007ea,Kruczenski:2008zk}. The background solution has a $AdS_2$ worldsheet embedded in $AdS_5$. Rotations of the remaining $AdS_5$ coordinates give an $SO(3)$ symmetry and since the string sits on a fixed point in $S^5$ the solution also has $SO(5)$ invariance.

It turns out that the 3 fluctuations in $AdS_5$ have $m^2=2$ (in units of the $AdS_2$ radius) while those coming from the 5-sphere are massless. These masses correspond to $SL(2,\mathds{R})$ quantum numbers $h=2$ and $h=1$, respectively. Thus, the bosonic spectrum respects the $SO(3)\times SO(5)$ symmetry of the solution. The fermionic fluctuations are described by eight real degrees of freedom which can be combined into four real 2d fermions. These fields have masses $|m|=1$ and transform in the fundamental of $SU(2)$ and the $\mathbf{4}$ of $SO(5)$. In terms of their $SL(2,\mathds{R})\times SO(3)\times SO(5)$ representations, the complete spectrum of excitations of the fundamental string dual to the BPS Wilson loop is given by
\begin{empheq}{align}
    \left(1,0,\mathbf{5}\right)\oplus\left(3/2,1/2,\mathbf{4}\right)\oplus\left(2,1,\mathbf{1}\right)
\end{empheq}
This is precisely the $\mathbf{j}=\mathbf{0}$ ultra-short multiplet of $OSp(4^*|4)$. Notice that this supersymmetric structure is in agreement with \cite{Drukker:2000ep}, where the authors argued that the fluctuations formed an $\mathcal{N}=8$ multiplet in two dimensions.
\subsection{D5 Brane}
In this section we discuss the spectrum of excitations of the D5 brane configuration corresponding to the BPS Wilson loop in the antisymmetric representation. We have to begin with a disclaimer stating that the results in this section are somehow speculative as a lot less is known in this case compared to the other two configurations discussed in this paper. We will defer the complete systematic discussion of this configuration to the future.

The classical configuration has been established to be a D5 brane in $AdS_5\times S^5$ with flux in its worldvolume. The induced geometry is $AdS_2\times S^4$, that is, the D5 brane wraps the $S^4$ inside the $S^5$.  Writing the $AdS_5\times S^5$  metric as
\begin{empheq}{align}
    ds^2&=L^2 \left(\cosh^2(u) ds_H^2+\sinh^2(u)d\Omega_2^2 + du^2 +d\theta^2 +\sin^2(\theta)d\Omega_4^2\right),
\end{empheq}
the solution is determined by the angle $\theta$ at which the brane sits (similar to equation (\ref{u0value}) for the D3 brane configuration). Namely \cite{Camino:2001at,Yamaguchi:2006tq},
\begin{empheq}{alignat=2}
    \theta&=\theta_k,
    &
    k&=\frac{2N}{\pi}\left(\frac12 \theta_k-\frac14 \sin 2\theta_k\right),
    \\\nonumber
    2\pi\alpha'F&=iL^2\cos(\theta_k)e^{\underline{0}}\wedge e^{\underline{1}},
\end{empheq}
where $k$ is the fundamental string charge on the brane.

Part of the bosonic spectrum of the D5 was obtained in \cite{Camino:2001at}, which considered the following fluctuations:
\begin{empheq}{alignat=2}
    \theta&=\theta_k+\xi,
    &\qquad
    F_{0r}&=\cos(\theta_k)+f.
\end{empheq}
Basically, the goal of the calculation in \cite{Camino:2001at} was to show that this configuration is stable. This can be achieved by considering only these excitations. However, for our goal, a full investigation along the lines of the one presented here for the D3 configuration is needed. At the risk of erring with our speculation, we proceed to organize the spectrum of excitations of the D5 brane into supermultiplets of $OSp(4^*|4)$.

The results for the two excitations considered in \cite{Camino:2001at} are (after the appropriate diagonalization):
\be\label{D5Alfonso}
m_l^2 =
\begin{cases}
(l+3)(l+4)& \text{for}\,\, l=0,1,\ldots \\
l(l-1)& \text{for}\,\, l=1,2, \ldots
\end{cases}
\ee
Here $l$ is related to the eigenvalue $l(l+2)$ of the spherical harmonic on $S^4$; we think of it as the $SO(5)$ quantum number. It determines the $SL(2,\mathds{R})$ quantum number via the formulas for conformal dimensions used in section \ref{Supersymmetry}.

Given that we do not have full information about the structure of the spectrum we will have to make some guesses. The representations involved in the full classification of the spectrum are different from the ones we presented in section \ref{Supersymmetry}. The crucial difference is that now we need to allow for higher representations of $SO(5)$. This forces us to switch to a more appropriate notation which is different from the one used until now. We find that the relevant representations can be summarized in table \ref{Table: D5 Representation} (see tables III and  IV of \cite{Gunaydin1991}):
\begin{table}
  \centering
  \begin{tabular}{|l|l|}
    \hline
    State & $\left(h,l\right)\times\left(m_1,m_2\right)$ \\
    \hline
    $\left|0\right\rangle$ & $(f,0)\times\left(0,f\right)$ \\
    \hline
    $\left|{\tiny\Yvcentermath1\yng(1)},{\tiny\Yvcentermath1\yng(1)}\right\rangle$ & $(f+\frac{1}{2},\frac{1}{2})\times\left(1,f-1\right)$ \\
    \hline
    $\left|{\tiny\Yvcentermath1\yng(2)},{\tiny\Yvcentermath1\yng(1,1)}\right\rangle$ & $(f+1,1)\times\left(0,f-1\right)$ \\
    \hline
    $\left|{\tiny\Yvcentermath1\yng(1,1)},{\tiny\Yvcentermath1\yng(2)}\right\rangle$ & $(f+1,0)\times\left(2,f-2\right)$ \\
    \hline
    $\left|{\tiny\Yvcentermath1\yng(2,1)},{\tiny\Yvcentermath1\yng(2,1)}\right\rangle$ & $(f+\frac{3}{2},\frac{1}{2})\times\left(1,f-2\right)$ \\
    \hline
    $\left|{\tiny\Yvcentermath1\yng(2,2)},{\tiny\Yvcentermath1\yng(2,2)}\right\rangle$ & $(f+2,0)\times\left(0,f-2\right)$ \\
    \hline
  \end{tabular}
  \caption{Short multiplets of $OSp(4^*|4)$ for $f\geq2$. When $f=1$ we get the ultrashort multiplet $\mathbf{0}$ introduced in section \ref{Supersymmetry}, which includes only the first 3 states.}\label{Table: D5 Representation}
\end{table}

In table \ref{Table: D5 Representation}, the first column represents the $U(2)\times U(2)\subset SO(4^*)\times USp(4)$ Young Tableaux and the second column gives the irreducible representations of $SO^*(4)\times SO(5)$. The $SO(5)$ representations are designated by their Dynkin labels. An important simplification is that we need to consider only singlets and triplets with respect $SO(3)$ for the the bosons; this was the role played by $SO(5)$ in the previous section for the D3 brane.

The lowest lying excitations quoted in (\ref{D5Alfonso}) correspond to
\begin{empheq}{alignat=3}
    h&=1,
    &\qquad
    \text{and}
    &\qquad
    h&=4.
\end{empheq}
With this information, and knowing that both states are singlets of $SO(3)$, we identify them as follows. The state with $h=4$ corresponds to the last state in table \ref{Table: D5 Representation} for $f=2$. The state with $h=1$ is identified with the first state of the ultrashort multiplet $f=1$. With these two modes identified, we move on to the excitations in $AdS$. These are relatively easy to recognize as they form a triplet of $SO(3)$. From the table we see that there is a unique choice, namely the third state $\left(f+1,1\right)\times\left(0,f-1\right)$. This leaves only one bosonic state in table \ref{Table: D5 Representation}, which must then correspond to the remaining gauge field fluctuations.

Perhaps a more careful way of phrasing this situation is to say (following the explicit calculation of \cite{Camino:2001at}) that there are linear combinations of excitations that fit into the multiplet structure described above. We make a prediction for the value of the $SL(2,\mathbb{R})$ quantum numbers or, equivalently, for the masses of the corresponding linear combinations. With some abuse of notation we will call these excitations triplet and singlets. We predict (in units of the $AdS$ radius):
\begin{empheq}{alignat=3}
    h_{triplet}&=f+1
    &\qquad
    &\longrightarrow
    &\qquad
    m^2_{triplet}&=f(f+1),
    \\\nonumber
    h_{singlet 1}&=f+2
    &\qquad
    &\longrightarrow
    &\qquad
    m^2_{singlet 1}&=(f+1)(f+2),
    \\\nonumber
    h_{singlet 2}&=f
    &\qquad
    &\longrightarrow
    &\qquad
    m^2_{singlet 2}&=(f-1)f
    \\\nonumber
    h_{singlet 3}&=f+1
    &\qquad
    &\longrightarrow
    &\qquad\nonumber
    m^2_{singlet 3}&=f(f+1).
\end{empheq}
It would be satisfying to verify this by explicit calculation.
\section{Conclusions}\label{Sec:conclusions}
In this paper we have tackled the question of excitations of classical configurations describing the expectation value of supersymmetric Wilson loops in ${\cal N}=4$ supersymmetric Yang-Mills in various representations of $SU(N)$. Concretely, we considered a fundamental string, a D3 brane and a D5 brane corresponding to the fundamental, symmetric and antisymmetric representations, respectively. We computed the spectrum of the D3 brane configuration explicitly. Our treatment of the fermionic excitations was exhaustive and we found interesting new properties that were not observed in previous analysis of the fermionic sector. Indeed, most of the analysis of fermions present in the literature has been rather indirect, relying on supersymmetry of the bosonic sector. We found, in particular, that the fermions obtained are massless in the worldvolume of the D3 brane, that is, on $AdS_2\times S^2$. This fact defies the naive expectation that a background with RR fluxes yields mass terms for the fermionic excitations. We basically witnessed an interesting cancelation between the would be mass term coming from the flux and the contribution of the extrinsic curvature. The results of the direct analysis fit precisely with the structure of supermultiplets of $OSp(4^*|4)$. After finding such harmonious picture for the D3 brane we moved on to the other two classical configurations that complete the holographic description of half BPS Wilson loops in ${\cal N}=4$ SYM. Namely, the fundamental string and the D5 brane with flux in its worldvolume. The spectrum of excitations of the fundamental string has been known for a long time but we have found its natural home in the ultrashort supermultiplet of $OSp(4^*|4)$. Much less is known about the spectrum of excitations of the D5 brane but we presented an educated guess of how to organize the spectrum based on the existing literature.

We will finish this section by mentioning a set of problems that we consider worth pursuing and will help to clarify some aspects of this beautiful duality between string theory configurations and expectation values of Wilson loop operators in ${\cal N}=4$ SYM. We list them in increasing speculative order:

\begin{itemize}
    \item It would be interesting to complete the calculation of excitations of the D5 brane. The interesting results of  \cite{Camino:2001at} demonstrate that the structure is more intricate than in the D3 brane case. The reorganization of the $SO(5)$ quantum numbers to yield a natural $SL(2,\mathds{R})$ structure for the masses of the fluctuations is quite nontrivial and requires the cooperation of various terms in the action. We hope to see this explicitly and believe that it might open the window to more general structures in AdS/CFT in the presence of D-branes with worldvolume fluxes. One could aim for a general result along the lines of the the one presented in \cite{Karch:2005ms} where it was shown that, under some general conditions, all the DBI excitations are simply classified by scalars in $AdS$.
    \\
    \item A natural next step for the work presented here is the computation of the one-loop determinants due to these fluctuations.  Knowing the one-loop determinant is equivalent to computing the first quantum correction to the expectation value of the half BPS Wilson loops. Clearly, such computation opens the door for comparison with other methods and exact results \cite{Pestun:2007rz}, and will provide further insight into the structure of AdS/CFT in supersymmetric setups. This is a very important computation and we plan to complete it in a separate publication. Let us just advance a few observations of what we glean from our experience here. It seems plausible to achieve a unified treatment of the straight line and the circular Wilson loop, whereby the only difference comes from global aspects of the $AdS_2$ space where the excitations live. One technical hurdle we anticipate is the fact that now we need to include the $SO(3)$ or $SO(5)$ quantum numbers when computing such determinants. Hopefully, the organization into supermultiplets achieved in this paper will serve as a guiding principle.
    \\
    \item Having understood the spectrum of excitations for 1/2 BPS holographic Wilson loops a natural question is whether the 1/4 BPS are amenable to a similar treatment. A clear starting point would be the solutions presented in \cite{Drukker:2006zk}.
    \\
    \item Given the prominent role that the matrix model has played in the context of general representations, it makes sense to expect that results similar to those obtained here could be mirrored in the matrix model side. In particular, it is likely that the computation of the expectation value of the Wilson loops could be organized in terms of excitations that are ultimately classified by $OSp(4^*|4)$.
    \\
    \item One of our original motivations for the study of these configurations was the hope that they might uncover some sort of integrable structure similar to those arising in the context of BMN or spin chains. We did not directly succeeded but hold out some hope that this is possible. We are encouraged by interesting works showing the role of integrability for circular Wilson loops using the fundamental string, even in the context of phase transitions \cite{Zarembo:1999bu,Burrington:2010yb}.
    \\
    \item Finally, in this paper we did not discuss the field theory dual in any detail. This is a one-dimensional defect CFT and has been quoted in recent works as a model for interesting condensed matter phenomena related to quantum impurity \cite{Sachdev:2010uj,Mueck:2010ja}. In such a context, uncovering the precise role of the spectrum of excitations should lead to a deeper understanding of the interactions of the system.
\end{itemize}
\section*{Acknowledgements}
We are grateful to Diego Rodr\'\i{}guez-G\'omez for collaboration on the initial stages of the project and various insightful comments. We would also like to thank Heng-Yu Chen,  Mart\'\i{}n Kruczenski, Luca Martucci, Alfonso Ramallo, Vincent Rodgers, Gordon Semenoff, Kory Stiffler, Alin Tirziu, and Diego Trancanelli, for comments. This work is  partially supported by Department of Energy under grant DE-FG02-95ER40899 to the University of Michigan. A.F. was supported by Fulbright-CONICYT fellowship.
\newpage
\appendix
\section{Notation and Conventions}\label{Appendix: Notation}
We work with the $AdS_5\times S^5$ metric
\begin{empheq}{align}
    ds^2&=ds^2_{AdS_5}+L^2d\Omega_5^2,
\end{empheq}
The $AdS_5$ metric is written as a $AdS_2\times S^2$ fibration
\begin{empheq}{align}
    ds^2_{AdS_5}&=L^2\left(\cosh^2(u)ds^2_H+\sinh^2(u)d\Omega_2^2+du^2\right).
\end{empheq}
where $ds^2_H$ is the line element of unit $AdS_2$.

Ten-dimensional indices are denoted by $m,\,n,\ldots$. We use $\mu,\,\nu,\ldots$ for indices along $AdS_2$, while indices on $S^2$ are labeled by $i,\,j,\ldots$. Directions transverse to $AdS_2\times S^2$ are denoted by $\hat{m}=(4,\hat{i})$, where $\hat{i}=5,6,7,8,9$ correspond to $S^5$. For worldvolume indices we use $\alpha,\,\beta,\ldots$. All tangent space indices are underlined.

We reserve the notation $e^{\underline{\mu}}$ for the unit $AdS_2$ vielbein and $e^{\underline{i}}$ for the unit $S^2$ vielbein. Similarly, the unit $S^5$ vielbein is denoted by $e^{\underline{\hat{i}}}$. The corresponding spin connections are $\omega^{\underline{\mu\nu}}$, $\omega^{\underline{ij}}$ and $\omega^{\underline{\hat{i}\hat{j}}}$, respectively. The $AdS_5\times S^5$ metric is labeled by $G_{mn}$ with vielbein $E^{\underline{m}}$ and spin connection $\Omega^{\underline{mn}}$. Then,
\begin{empheq}{alignat=4}
    E^{\underline{\mu}}&=L\cosh(u)e^{\underline{\mu}},
    &\qquad
    E^{\underline{i}}&=L\sinh(u)e^{\underline{\mu}},
    &\qquad
    E^{\underline{4}}&=Ldu,
    &\qquad
    E^{\underline{\hat{i}}}&=Le^{\underline{\hat{i}}},
\end{empheq}
and
\begin{empheq}{alignat=5}
    \Omega^{\underline{\mu\nu}}&=\omega^{\underline{\mu\nu}},
    &\qquad
    \Omega^{\underline{\mu4}}&=\sinh(u)e^{\underline{\mu}},
    &\qquad
    \Omega^{\underline{ij}}&=\omega^{\underline{ij}},
    &\qquad
    \Omega^{\underline{i4}}&=\cosh(u)e^{\underline{i}},
    &\qquad
    \Omega^{\underline{\hat{i}\hat{j}}}&=\omega^{\underline{\hat{i}\hat{j}}}.
\end{empheq}

The induced metric is $g_{\alpha\beta}\equiv P[G]_{\alpha\beta}$, where $P[\;]$ denotes the pullback to the worldvolume. The corresponding spin connection is $w^{\underline{\alpha\beta}}$. We do not make explicit reference to the vielbein for this metric.

The deformed metric is denoted by $\hat{g}_{\alpha\beta}$, its vielbein by $\hat{e}^{\underline{\alpha}}$ and its spin connection by $\hat{w}^{\underline{\alpha\beta}}$.
\section{Explicit Calculations}\label{Appendix: Explicit Calculations}
In this appendix we show the details of the calculation leading to the actions \eqref{Quadratic Bosonic Action} and \eqref{Quadratic Fermionic Action}. For simplicity we will discuss the infinite line in Lorentzian signature. This means that we use \eqref{AdS_2 Half-Plane Lorentzian} for the metric on $AdS_2$ and the background gauge field \eqref{Gauge Field Lorentzian}. We also set $2\pi\alpha'=1$. The calculation for the circular loop is completely analogous but must be done in Euclidean signature.
\subsection{Bosons}
In Lorentzian signature the bosonic action is
\begin{empheq}{align}
    S_B&=-T_{D3}\int d^4\sigma\,\sqrt{-\textrm{det}\left(M\right)}+T_{D3}\int P[C_4],
\end{empheq}
where
\begin{empheq}{align}
    M_{\alpha\beta}&=g_{\alpha\beta}+F_{\alpha\beta}.
\end{empheq}
We want to expand this action to second order in the fluctuations (in the main text we called $\delta u=\phi^4$ and $\delta\theta^{\hat{i}}=\phi^{\hat{i}}$)
\begin{empheq}{alignat=3}
    u&=u_k+\delta u,
    &\quad
    \theta^{\hat{i}}&=\theta^{\hat{i}}_0+\delta\theta^{\hat{i}},
    &\quad  F=\frac{L^2\cosh(u_k)}{r^2}dt\wedge dr+f.
\end{empheq}
Since they are decoupled from the rest, we will ignore the fluctuations $\delta\theta^{\hat{i}}$.

In matrix form, the classical induced metric and gauge field read
\begin{empheq}{align}
    \overline{g}_{\alpha\beta}&=L^2\left(
                          \begin{array}{cccc}
                            {\displaystyle-\frac{\cosh^2(u_k)}{r^2}} & 0 & 0 & 0 \\
                            0 & {\displaystyle\frac{\cosh^2(u_k)}{r^2}} & 0 & 0 \\
                            0 & 0 & \sinh^2(u_k) & 0 \\
                            0 & 0 & 0 & \sinh^2(u_k)\sin^2\theta \\
                          \end{array}
                        \right),
    \\
    \overline{F}_{\alpha\beta}&=\frac{L^2\cosh(u_k)}{r^2}\left(
                                                                       \begin{array}{cccc}
                                                                         0 & 1 & 0 & 0 \\
                                                                         -1 & 0 & 0 & 0 \\
                                                                         0 & 0 & 0 & 0 \\
                                                                         0 & 0 & 0 & 0 \\
                                                                       \end{array}
                                                                     \right).
\end{empheq}
Thus, the tensor $M_{\alpha\beta}$ and its inverse $M^{\alpha\beta}$ evaluated on the solution are
\begin{empheq}{align}
    \overline{M}_{\alpha\beta}&=L^2\left(
                          \begin{array}{cccc}
                            {\displaystyle-\frac{\cosh^2(u_k)}{r^2}} & {\displaystyle\frac{\cosh(u_k)}{r^2}} & 0 & 0 \\
                            {\displaystyle-\frac{\cosh(u_k)}{r^2}} & {\displaystyle\frac{\cosh^2(u_k)}{r^2}} & 0 & 0 \\
                            0 & 0 & \sinh^2(u_k) & 0 \\
                            0 & 0 & 0 & \sinh^2(u_k)\sin^2\theta \\
                          \end{array}
                        \right)
    \\
    \overline{M}^{\alpha\beta}&=\frac{1}{L^2\sinh^2(u_k)}\left(
                                                \begin{array}{cccc}
                                                  -r^2 & {\displaystyle\frac{r^2}{\cosh(u_k)}} & 0 & 0 \\
                                                  {\displaystyle-\frac{r^2}{\cosh(u_k)}} & r^2 & 0 & 0 \\
                                                  0 & 0 & 1 & 0 \\
                                                  0 & 0 & 0 & {\displaystyle\frac{1}{\sin^2\theta}} \\
                                                \end{array}
                                              \right).
\end{empheq}
At this point we can already see the emergence of the deformed geometry. The diagonal part of $\overline{M}^{\alpha\beta}$ is precisely the inverse of \eqref{Deformed Geometry}.

Now, expanding the induced metric $g_{\alpha\beta}$ we get
\begin{empheq}{align}
    ds^2&=L^2\left(\cosh^2(u_k)ds^2_H+\sinh(u_k)d\Omega_2^2\right)
    \\\nonumber
    &+2L^2\sinh(u_k)\cosh(u_k)\left(ds^2_H+d\Omega_2^2\right)\delta u
    \\\nonumber
    &+L^2\left(\cosh^2(u_k)+\sinh^2(u_k)\right)\left(ds^2_H+d\Omega_2^2\right)\delta u^2+L^2d\delta u^2.
\end{empheq}
The first line is the induced metric on the classical solution. The second and third lines are the linear and quadratic fluctuations, respectively. Define
\begin{empheq}{align}
    l_{\alpha\beta}&=2L^2\sinh(u_k)\cosh(u_k)\left(
                                               \begin{array}{cccc}
                                                 {\displaystyle-\frac{1}{r^2}} & 0 & 0 & 0 \\
                                                 0 & {\displaystyle\frac{1}{r^2}} & 0 & 0 \\
                                                 0 & 0 & 1 & 0 \\
                                                 0 & 0 & 0 & \sin^2\theta \\
                                               \end{array}
                                             \right)\delta u,
    \\
    q_{\alpha\beta}&=L^2\left(\cosh^2(u_k)+\sinh^2(u_k)\right)\left(
                                                                \begin{array}{cccc}
                                                                  {\displaystyle-\frac{1}{r^2}} & 0 & 0 & 0 \\
                                                                  0 & {\displaystyle\frac{1}{r^2}} & 0 & 0 \\
                                                                  0 & 0 & 1 & 0 \\
                                                                  0 & 0 & 0 & \sin^2\theta \\
                                                                \end{array}
                                                              \right)\delta u^2,
    \\
    d_{\alpha\beta}&=L^2\partial_{\alpha}\delta u\partial_{\beta}\delta u.
\end{empheq}
Then,
\begin{empheq}{align}
    M_{\alpha\beta}&=\overline{M}_{\alpha\beta}+l_{\alpha\beta}+q_{\alpha\beta}+d_{\alpha\beta}+f_{\alpha\beta}.
\end{empheq}
Raising indices with $\overline{M}^{\alpha\beta}$ we find
\begin{empheq}{align}
    l^{\alpha}_{\phantom{\alpha}\beta}&=2\coth(u_k)\left(
                                                     \begin{array}{cccc}
                                                       1 & {\displaystyle\frac{1}{\cosh(u_k)}} & 0 & 0 \\
                                                       {\displaystyle\frac{1}{\cosh(u_k)}} & 1 & 0 & 0 \\
                                                       0 & 0 & 1 & 0 \\
                                                       0 & 0 & 0 & 1 \\
                                                     \end{array}
                                                   \right)\delta u,
    \\
    q^{\alpha}_{\phantom{\alpha}\alpha}&=4\left(1+\coth^2(u_k)\right)\delta u^2,
    \\
    d^{\alpha}_{\phantom{\alpha}\alpha}&=\frac{1}{\sinh^2(u_k)}\left(-r^2\left(\partial_t\delta u\right)^2+r^2\left(\partial_r\delta u\right)^2+\left(\partial_{\theta}\delta u\right)^2+\frac{1}{\sin^2\theta}\left(\partial_{\phi}\delta u\right)^2\right),
    \\
    f^{\alpha}_{\phantom{\alpha}\beta}&=\frac{1}{L^2\sinh^2(u_k)}\left(
                                                                   \begin{array}{cccc}
                                                                     {-\frac{r^2f_{tr}}{\cosh(u_k)}} & -r^2f_{tr} & {-r^2f_{t\theta}+\frac{r^2f_{r\theta}}{\cosh(u_k)}} & {-r^2f_{t\phi}+\frac{r^2f_{r\phi}}{\cosh(u_k)}} \\
                                                                     {-r^2f_{tr}} & {-\frac{r^2f_{tr}}{\cosh(u_k)}} & {-\frac{r^2f_{t\theta}}{\cosh(u_k)}+r^2f_{r\theta}} & {-\frac{r^2f_{t\phi}}{\cosh(u_k)}+r^2f_{r\phi}} \\
                                                                     -f_{t\theta} & -f_{r\theta} & 0 & f_{\theta\phi} \\
                                                                     {-\frac{f_{t\phi}}{\sin^2\theta}} & {-\frac{f_{r\phi}}{\sin^2\theta}} & {-\frac{f_{\theta\phi}}{\sin^2\theta}} & 0 \\
                                                                   \end{array}
                                                                 \right),
\end{empheq}
where we have introduced the fluctuation of the gauge field in matrix form.

In order to expand the determinant of $M_{\alpha\beta}$ we make use of the expression
\begin{empheq}{align}
    \sqrt{\textrm{det}\left(1+A\right)}&=1+\frac{1}{2}\textrm{Tr}\left(A\right)+\frac{1}{8}\textrm{Tr}^2\left(A\right)-\frac{1}{4}\textrm{Tr}\left(A^2\right)+O(A^3).
\end{empheq}
The linear terms coming from this expansion are
\begin{empheq}{align}
    \frac{1}{2}l^{\alpha}_{\phantom{\alpha}\alpha}&=4\coth(u_k)\delta u,
    \\
    \frac{1}{2}f^{\alpha}_{\phantom{\alpha}\alpha}&=-\frac{r^2f_{tr}}{L^2\sinh^2(u_k)\cosh(u_k)}.
\end{empheq}
Similarly, the quadratic terms read
\begin{empheq}{align}
    \frac{1}{2}q^{\alpha}_{\phantom{\alpha}\alpha}&=2\left(1+\coth^2(u_k)\right)\delta u^2,
    \\
    \frac{1}{2}d^{\alpha}_{\phantom{\alpha}\alpha}&=\frac{1}{2\sinh^2(u_k)}\left(-r^2\left(\partial_t\delta u\right)^2+r^2\left(\partial_r\delta u\right)^2+\left(\partial_{\theta}\delta u\right)^2+\frac{1}{\sin^2\theta}\left(\partial_{\phi}\delta u\right)^2\right),
\end{empheq}
and
\begin{empheq}{align}
    \frac{1}{8}\left(l^{\alpha}_{\phantom{\alpha}\alpha}\right)^2-\frac{1}{4}l^{\alpha}_{\phantom{\alpha}\beta}l^{\beta}_{\phantom{\beta}\alpha}&=2\left(1+\coth^2(u_k)\right)\delta u^2,
    \\
    \frac{1}{8}\left(f^{\alpha}_{\phantom{\alpha}\alpha}\right)^2-\frac{1}{4}f^{\alpha}_{\phantom{\alpha}\beta}f^{\beta}_{\phantom{\beta}\alpha}&=\frac{1}{2L^4\sinh^4(u_k)}\left(-r^4f_{tr}^2-r^2f_{t\theta}^2-\frac{r^2f_{t\phi}^2}{\sin^2\theta}+r^2f_{r\theta}^2\right.
    \\\nonumber
    &\left.+\frac{r^2f_{r\phi}^2}{\sin^2\theta}+\frac{f_{\theta\phi}^2}{\sin^2\theta}\right),
    \\
    \frac{1}{4}l^{\alpha}_{\phantom{\alpha}\alpha}f^{\beta}_{\phantom{\beta}\beta}-\frac{1}{2}l^{\alpha}_{\phantom{\alpha}\beta}f^{\beta}_{\phantom{\beta}\alpha}&=0.
\end{empheq}
Notice that the mixing between the scalar and gauge field fluctuations vanishes. With this, the DBI contribution to the action is
\begin{empheq}{align}
    \sqrt{-\textrm{det}\left(M\right)}&=\sqrt{-\textrm{det}\left(\overline{M}\right)}\Bigg[1+4\coth(u_k)\delta u-\frac{r^2f_{tr}}{L^2\sinh^2(u_k)\cosh(u_k)}
    \\\nonumber
    &+\frac{1}{2\sinh^2(u_k)}\left(-r^2\left(\partial_t\delta u\right)^2+r^2\left(\partial_r\delta u\right)^2+\left(\partial_{\theta}\delta u\right)^2+\frac{1}{\sin^2\theta}\left(\partial_{\phi}\delta u\right)^2\right)
    \\\nonumber
    &+\frac{1}{2L^4\sinh^4(u_k)}\left(-r^4f_{tr}^2-r^2f_{t\theta}^2-\frac{r^2f_{t\phi}^2}{\sin^2\theta}+r^2f_{r\theta}^2+\frac{r^2f_{r\phi}^2}{\sin^2\theta}+\frac{f_{\theta\phi}^2}{\sin^2\theta}\right)
    \\\nonumber
    &+4\left(1+\coth^2(u_k)\right)\delta u^2\Bigg],
\end{empheq}
where
\begin{empheq}{align}
    \sqrt{-\textrm{det}\left(\overline{M}\right)}&=\frac{L^4\cosh(u_k)\sinh^3(u_k)\sin\theta}{r^2}.
\end{empheq}
The term linear in the gauge field is a total derivative.

Moving on to the WZ term we have
\begin{empheq}{align}
    C_4&=\frac{4L^4f(u)\sin\theta}{r^2}dt\wedge dr\wedge d\theta\wedge d\phi,
\end{empheq}
where
\begin{empheq}{align}
    f(u)&=\frac{1}{8}\sinh(u)\cosh(u)\left(\cosh^2(u)+\sinh^2(u)\right)-\frac{u}{8}.
\end{empheq}
Expanding to second order we get
\begin{empheq}{align}
    f(u_k+\delta u)&=\frac{1}{8}\sinh(u_k)\cosh(u_k)\left(\cosh^2(u_k)+\sinh^2(u_k)\right)-\frac{u_k}{8}
    \\\nonumber
    &+\cosh^2(u_k)\sinh^2(u_k)\delta u
    \\\nonumber
    &+\sinh(u_k)\cosh(u_k)\left(\cosh^2(u_k)+\sinh^2(u_k)\right)\delta u^2.
\end{empheq}
Then,
\begin{empheq}{align}
    P[C_4]&=\sqrt{-\textrm{det}\left(\overline{M}_{\alpha\beta}\right)}\Bigg[\frac{1}{2}\left(1+\coth^2(u_k)\right)-\frac{u_k}{2\cosh(u_k)\sinh^3(u_k)}
    \\\nonumber
    &+4\coth(u_k)\delta u+4\left(1+\coth^2(u_k)\right)\delta u^2\Bigg]dt\wedge dr\wedge d\theta\wedge d\phi.
\end{empheq}

Putting everything together we get
\begin{empheq}{align}
    S_B&=\int d^4\sigma\,\sqrt{\textrm{det}\left(\overline{M}_{\alpha\beta}\right)}\Bigg[\frac{1}{2\sinh^2(u_k)}-\frac{u_k}{2\cosh(u_k)\sinh^3(u_k)}
    \\\nonumber
    &-\frac{1}{2\sinh^2(u_k)}\left(-r^2\left(\partial_t\delta u\right)^2+r^2\left(\partial_r\delta u\right)^2+\left(\partial_{\theta}\delta u\right)^2+\frac{1}{\sin^2\theta}\left(\partial_{\phi}\delta u\right)^2\right)
    \\\nonumber
    &-\frac{1}{2L^4\sinh^4(u_k)}\left(-r^4f_{tr}^2-r^2f_{t\theta}^2-\frac{r^2f_{t\phi}^2}{\sin^2\theta}+r^2f_{r\theta}^2+\frac{r^2f_{r\phi}^2}{\sin^2\theta}+\frac{f_{\theta\phi}^2}{\sin^2\theta}\right)\Bigg].
\end{empheq}
The first line is the on-shell action. As expected, the term linear in $\delta u$ cancels. Notice also that the mass terms coming from the DBI and WZ actions cancel out. Finally, introducing the deformed metric
\begin{empheq}{align}
    \hat{g}_{\alpha\beta}&=L^2\sinh^2(u_k)\left(
                                            \begin{array}{cccc}
                                              {\displaystyle-\frac{1}{r^2}} & 0 & 0 & 0 \\
                                              0 & {\displaystyle\frac{1}{r^2}} & 0 & 0 \\
                                              0 & 0 & 1 & 0 \\
                                              0 & 0 & 0 & \sin^2\theta \\
                                            \end{array}
                                          \right),
\end{empheq}
the quadratic term can be written as
\begin{empheq}{align}
    S_B^{(2)}&=-T_{D3}\coth(u_0)\int d^4\sigma\sqrt{-\hat{g}}\left(\frac{1}{2}L^2\hat{g}^{\alpha\beta}\partial_{\alpha}\delta u\partial_{\beta}\delta u+\frac{1}{4}\hat{g}^{\alpha\beta}\hat{g}^{\gamma\delta}f_{\alpha\gamma}f_{\beta\delta}\right).
\end{empheq}
\subsection{Fermions}
\subsubsection{Fermionic action}
We want to compute the fermionic Lagrangian
\begin{empheq}{align}
    \mathcal{L}_F&=\overline{\Theta}\left(1-\Gamma_{D3}\right)\tilde{M}^{\alpha\beta}\Gamma_{\beta}D_{\alpha}\Theta
\end{empheq}
explicitly. First, when evaluated on the classical solution, the tensor $\tilde{M}_{\alpha\beta}$ defined in \eqref{Definition of tilde M} reads
\begin{empheq}{align}
    \tilde{M}_{\alpha\beta}&=L^2\left(
                                  \begin{array}{cccc}
                                    {\displaystyle-\frac{\cosh^2(u_k)}{r^2}} & {\displaystyle\frac{\cosh(u_k)}{r^2}\tilde{\Gamma}} & 0 & 0 \\
                                    {\displaystyle-\frac{\cosh(u_k)}{r^2}\tilde{\Gamma}} & {\displaystyle\frac{\cosh^2(u_k)}{r^2}} & 0 & 0 \\
                                    0 & 0 & \sinh^2(u_k) & 0 \\
                                    0 & 0 & 0 & \sinh^2(u_k)\sin^2\theta \\
                                  \end{array}
                                \right),
\end{empheq}
and its inverse is
\begin{empheq}{align}\label{tilde M Inverse}
    \tilde{M}^{\alpha\beta}&=\frac{1}{L^2\sinh^2(u_k)}\left(
                                                        \begin{array}{cccc}
                                                          -r^2 & {\displaystyle\frac{r^2}{\cosh(u_k)}\tilde{\Gamma}} & 0 & 0 \\
                                                          {-\displaystyle\frac{r^2}{\cosh(u_k)}\tilde{\Gamma}} & r^2 & 0 & 0 \\
                                                          0 & 0 & 1 &  \\
                                                          0 & 0 & 0 & {\displaystyle\frac{1}{\sin^2\theta}} \\
                                                        \end{array}
                                                      \right).
\end{empheq}
As already noted in the bosonic calculation, the diagonal part of $\tilde{M}^{\alpha\beta}$ is the inverse of the deformed metric \eqref{Definition of Deformed Metric}.

Now, the pullback of the 10d Dirac matrices is
\begin{empheq}{alignat=4}
    \Gamma_t&=\frac{L\cosh(u_k)}{r}\Gamma_{\underline{0}},
    &\quad
    \Gamma_r&=\frac{L\cosh(u_k)}{r}\Gamma_{\underline{1}},
    \\\nonumber
    \Gamma_{\theta}&=L\sinh(u_k)\Gamma_{\underline{2}},
    &\quad
    \Gamma_{\phi}&=L\sinh(u_k)\sin\theta\Gamma_{\underline{3}}.
\end{empheq}
With these expressions we find
\begin{empheq}{align}
    \tilde{M}^{\alpha\beta}\Gamma_{\beta}D_{\alpha}&=\frac{1}{L\sinh(u_k)}\left[-re^{2R\tilde{\Gamma}}\Gamma_0D_t+re^{2R\tilde{\Gamma}}\Gamma_1D_r+\Gamma_2D_{\theta}+\frac{1}{\sin\theta}\Gamma_3D_{\phi}\right],
\end{empheq}
where the antisymmetric part of \eqref{tilde M Inverse} has been absorbed in
\begin{empheq}{align}
    R&=-\frac{1}{2}\sinh^{-1}\left(\frac{1}{\sinh(u_k)}\right)\Gamma_{01}.
\end{empheq}
Using the vielbein
\begin{empheq}{alignat=2}
    \hat{e}^{\underline{0}}&=\frac{L\sinh(u_k)}{r}dt,
    &\qquad
    \hat{e}^{\underline{1}}&=\frac{L\sinh(u_k)}{r}dr,
    \\\nonumber
    \hat{e}^{\underline{2}}&=L\sinh(u_k)d\theta,
    &\qquad
    \hat{e}^{\underline{3}}&=L\sinh(u_k)\sin\theta d\phi,
\end{empheq}
for the deformed metric $\eqref{Deformed Geometry}$, we can further simplify this to
\begin{empheq}{align}
    \tilde{M}^{\alpha\beta}\Gamma_{\beta}D_{\alpha}&=e^{R\tilde{\Gamma}}\left[\hat{\Gamma}^{\alpha}e^{R\tilde{\Gamma}}D_{\alpha}e^{-R\tilde{\Gamma}}\right]e^{R\tilde{\Gamma}},
\end{empheq}
where $\hat{\Gamma}_{\alpha}=\hat{e}^{\underline{\alpha}}_{\phantom{\underline{\alpha}}\alpha}\Gamma_{\underline{\alpha}}$ and $\hat{\Gamma}^{\alpha}=\hat{g}^{\alpha\beta}\hat{\Gamma}_{\beta}$.

Next, by looking at the spin connection written in Appendix \ref{Appendix: Notation}, we find that the pullback of the 10d covariant derivative is
\begin{empheq}{align}
    \nabla_{\alpha}d\sigma^{\alpha}&=d+\frac{1}{4}\omega^{\underline{\mu\nu}}\Gamma_{\underline{\mu\nu}}+\frac{1}{4}\omega^{\underline{ij}}\Gamma_{\underline{ij}}+\frac{1}{2}\sinh(u_k)e^{\underline{\mu}}\Gamma_{\underline{\mu4}}+\frac{1}{2}\cosh(u_k)e^{\underline{i}}\Gamma_{\underline{i4}}.
\end{empheq}
The first three terms form the covariant derivative with respect to the deformed worldvolume geometry \eqref{Deformed Geometry}, which we call $\hat{\nabla}$\footnote{Referring to the formulas in Appendix \ref{Appendix: Notation}, it is easy to see that $w^{\underline{\alpha\beta}}=\hat{w}^{\underline{\alpha\beta}}=P[\Omega^{\underline{\alpha\beta}}]$}. The other two terms, which come from the extrinsic curvature of the worldvolume, are potential mass terms for the fermionic field $\Theta$. We find that
\begin{empheq}{align}
    e^{R\tilde{\Gamma}}\nabla_{\alpha}d\sigma^{\alpha} e^{-R\tilde{\Gamma}}&=\hat{\nabla}_{\alpha}d\sigma^{\alpha}+\frac{1}{2}\sinh(u_k)e^{\underline{\mu}}\Gamma_{\underline{\mu4}}e^{-2R\tilde{\Gamma}}+\frac{1}{2}\cosh(u_k)e^{\underline{i}}\Gamma_{\underline{i4}}.
\end{empheq}
Then,
\begin{empheq}{align}
    \tilde{M}^{\alpha\beta}\Gamma_{\beta}\nabla_{\alpha}&=e^{R\tilde{\Gamma}}\left[\hat{\Gamma}^{\alpha}\hat{\nabla}_{\alpha}+\frac{1}{L\sinh(u_k)}\Gamma_{\underline{4}}\left(\sinh(u_k)e^{-2R\tilde{\Gamma}}+\cosh(u_k)\right)\right]e^{R\tilde{\Gamma}}.
\end{empheq}

The other term in the IIB covariant derivative \eqref{IIB Covariant Derivative 10d} comes from the RR 5-form. A short calculation shows
\begin{empheq}{align}
    \slashed{F}_5\Gamma_{\alpha}d\sigma^{\alpha}\otimes\left(i\sigma_2\right)&=-4\Gamma_{\underline{01234}}\otimes\left(i\sigma_2\right)\left(\cosh(u_k)e^{\underline{\mu}}\Gamma_{\underline{\mu}}+\sinh(u_k)e^{\underline{i}}\Gamma_{\underline{i}}\right)\left(1+\Gamma^{11}\right).
\end{empheq}
Since the spinor $\Theta$ satisfies $\Gamma^{11}\Theta=\Theta$ we can replace $\Gamma^{11}=1$ in this expression. Thus,
\begin{empheq}{align}
    e^{R\tilde{\Gamma}}\slashed{F}_5\Gamma_{\alpha}d\sigma^{\alpha}\otimes\left(i\sigma_2\right)e^{-R\tilde{\Gamma}}&=-8\Gamma_{\underline{01234}}\otimes\left(i\sigma_2\right)\left(\cosh(u_k)e^{\underline{\mu}}\Gamma_{\underline{\mu}}+\sinh(u_k)e^{\underline{i}}\Gamma_{\underline{i}}e^{-2R\tilde{\Gamma}}\right),
\end{empheq}
and we get
\begin{empheq}{align}
    \tilde{M}^{\alpha\beta}\Gamma_{\beta}\slashed{F}_5\Gamma_{\alpha}\otimes\left(i\sigma_2\right)&=-\frac{16e^{R\tilde{\Gamma}}}{L\sinh(u_k)}\Gamma_{\underline{01234}}\otimes\left(i\sigma_2\right)\left(\cosh(u_k)+\sinh(u_k)e^{-2R\tilde{\Gamma}}\right)e^{R\tilde{\Gamma}}.
\end{empheq}
Putting these results together we see that
\begin{empheq}{align}
    \tilde{M}^{\alpha\beta}\Gamma_{\beta}D_{\alpha}&=e^{R\tilde{\Gamma}}\left[\hat{\Gamma}^{\alpha}\hat{\nabla}_{\alpha}+\frac{\left(1-\Gamma^{(0)}_{D3}\right)}{L\sinh(u_k)}\Gamma_{\underline{4}}\left(\sinh(u_k)e^{-2R\tilde{\Gamma}}+\cosh(u_k)\right)\right]e^{R\tilde{\Gamma}},
\end{empheq}
where
\begin{empheq}{align}
    \Gamma^{(0)}_{D3}&=\Gamma_{0123}\otimes\left(i\sigma_2\right).
\end{empheq}

Now we look at the $\kappa$-symmetry projector $\Gamma_{D3}$ as defined in \cite{Martucci:2005rb}:
\begin{empheq}{align}\label{Definition of Gamma_D3}
    \Gamma_{D3}&=-\frac{\epsilon^{\alpha_1\alpha_2\alpha_3\alpha_4}\Gamma_{\alpha_1\alpha_2\alpha_3\alpha_4}}{(p+1)!\sqrt{\textrm{det}\left(g+F\right)}}\otimes\left(i\sigma_2\right)\times\sum_q\frac{\left(\sigma_3\right)^q}{q!2^q}\otimes\Gamma^{\alpha_1\cdots\alpha_{2q}}F_{\alpha_1\alpha_2}\cdots F_{\alpha_{2q-1}\alpha_{2q}},
\end{empheq}
where $\epsilon^{0123}=1$. In our case we find
\begin{empheq}{align}
    \Gamma_{D3}&=-\Gamma^{(0)}_{D3}\left(\coth(u_k)-\frac{1}{\sinh(u_k)}\Gamma_{01}\otimes\left(\sigma_3\right)\right).
\end{empheq}
Acting on the IIB spinor $\Theta$, this can be written as
\begin{empheq}{align}
    \overline{\Theta}\Gamma_{D3}&=-\overline{\Theta}e^{R\tilde{\Gamma}}\Gamma^{(0)}_{D3}e^{-R\tilde{\Gamma}}.
\end{empheq}

Using the above expressions, the fermionic D3 brane Lagrangian reads
\begin{empheq}{align}
    \mathcal{L}_F&=\overline{\Theta}e^{R\tilde{\Gamma}}\left(1+\Gamma^{(0)}_{D3}\right)\left[\hat{\Gamma}^{\alpha}\hat{\nabla}_{\alpha}+\frac{\left(1-\Gamma^{(0)}_{D3}\right)}{L\sinh(u_k)}\Gamma_{\underline{4}}\left(\sinh(u_k)e^{-2R\tilde{\Gamma}}+\cosh(u_k)\right)\right]e^{R\tilde{\Gamma}}\Theta.
\end{empheq}
Notice that $\left(\Gamma_{D3}^{(0)}\right)^2=1$. This implies that the second term vanishes, making the fermions massless. Redefining $e^{R\tilde{\Gamma}}\Theta=\Theta'\Rightarrow\overline{\Theta}e^{R\tilde{\Gamma}}=\overline{\Theta}'$ we get
\begin{empheq}{align}
    \mathcal{L}_F&=\overline{\Theta}'\left(1+\Gamma^{(0)}_{D3}\right)\hat{\Gamma}^{\alpha}\hat{\nabla}_{\alpha}\Theta'.
\end{empheq}
Finally, we fix the $\kappa$-symmetry by demanding
\begin{empheq}{align}
    \tilde{\Gamma}\Theta'&=\Theta'.
\end{empheq}
In other words, the lower component of the doublet $\Theta'$ is zero. This way we get the final form of the Lagrangian (dropping the primes):
\begin{empheq}{align}
    \overline{\Theta}\left(1-\Gamma_{D3}\right)\tilde{M}^{\alpha\beta}\Gamma_{\beta}D_{\alpha}\Theta&=\overline{\Theta}\hat{\Gamma}^{\alpha}\hat{\nabla}_{\alpha}\Theta.
\end{empheq}
\subsubsection{Dimensional reduction}
Let us do the dimensional reduction of the spinor field $\Theta$ from 10d down to 4d. First we choose a particular representation of 10d Dirac matrices:
\begin{empheq}{alignat=2}
    \Gamma_{\underline{\alpha}}&=\gamma_{\underline{\alpha}}\otimes\mathds{1},
    &\quad
    \Gamma_{\underline{\hat{m}}}&=\gamma_5\otimes\rho_{\underline{\hat{m}}},
\end{empheq}
where $\gamma_{\underline{\alpha}}$ and $\rho_{\underline{\hat{m}}}$ are $SO(3,1)$ and $SO(6)$ Dirac matrices, respectively. Also, $\gamma_5=i\gamma^{\underline{0123}}$. Then, the ten-dimensional chirality matrix is
\begin{empheq}{align}
    \Gamma^{11}&=-\gamma_5\otimes\rho_{10},
\end{empheq}
where $\rho_{10}=i\rho^{\underline{456789}}$. Furthermore, we use the intertwiners
\begin{empheq}{alignat=2}
    B_{(9,1)\pm}\Gamma_{\underline{m}}B_{(9,1)\pm}^{-1}&=\pm\Gamma_{\underline{m}}^*,
    &\qquad
    B_{(9,1)\pm}^T&=B_{(9,1)\pm},
    \\\nonumber
    B_{(3,1)\pm}\gamma_{\underline{\alpha}}B_{(3,1)\pm}^{-1}&=\pm\gamma_{\underline{\alpha}}^*,
    &\qquad
    B_{(3,1)\pm}^T&=\pm B_{(3,1)\pm},
    \\\nonumber
    B_{(6,0)\pm}\rho_{\underline{\hat{m}}}B_{(6,0)\pm}^{-1}&=\pm\rho_{\underline{\hat{m}}}^*,
    &\qquad
    B_{(6,0)\pm}^T&=\mp B_{(6,0)\pm}.
\end{empheq}
They also satisfy,
\begin{empheq}{alignat=3}
    B_{(9,1)\pm}\Gamma_{11}B_{(9,1)\pm}^{-1}&=\Gamma_{11}^*,
    &\qquad
    B_{(3,1)\pm}\gamma_{5}B_{(3,1)\pm}^{-1}&=-\gamma_{5}^*,
    &\qquad
    B_{(6,0)\pm}\rho_{10}B_{(6,0)\pm}^{-1}&=-\rho_{10}^*.
\end{empheq}
Using the above decomposition, we find that
\begin{empheq}{align}
    B_{(9,1)\pm}&=B_{(3,1)\pm}\otimes B_{(6,0)\mp}
\end{empheq}

We can build a representation of $SO(6)$ gamma matrices by taking products of three Pauli matrices. We choose
\begin{empheq}{alignat=3}
    \rho_{\underline{4}}&=\sigma_2\otimes\mathds{1}\otimes\mathds{1},
    &\quad
    \rho_{\underline{5}}&=\sigma_1\otimes\sigma_1\otimes\sigma_2,
    &\quad
    \rho_{\underline{6}}&=\sigma_3\otimes\mathds{1}\otimes\sigma_2,
    \\\nonumber
    \rho_{\underline{7}}&=\sigma_3\otimes\sigma_3\otimes\sigma_1,
    &\quad
    \rho_{\underline{8}}&=\sigma_1\otimes\sigma_3\otimes\mathds{1},
    &\quad
    \rho_{\underline{9}}&=\sigma_1\otimes\sigma_2\otimes\sigma_2.
\end{empheq}
Then,
\begin{empheq}{align}
    \rho_{10}&=-\sigma_3\otimes\sigma_3\otimes\sigma_3,
\end{empheq}
and
\begin{empheq}{alignat=3}
    B_{(6,0)-}&=\gamma_{\underline{456}}&=-i\mathds{1}\otimes\sigma_1\otimes\mathds{1},
    \\
    B_{(6,0)+}&=\gamma_{\underline{789}}&=i\sigma_3\otimes\sigma_2\otimes\sigma_3
\end{empheq}

Now, a general 10d spinor can be expanded as
\begin{empheq}{align}
    \Theta&=\sum_{\alpha_i}\Theta^{\alpha_1\alpha_2\alpha_3}\otimes\eta_{\alpha_1}\otimes\eta_{\alpha_2}\otimes\eta_{\alpha_3},
\end{empheq}
where $\Theta^{\alpha_1\alpha_2\alpha_3}$ are $SO(3,1)$ spinors and $\eta_{\alpha}$ are 2d spinors with $U(1)$ charge $\alpha$, i.e. $\sigma_3\eta_{\alpha}=\alpha\eta_{\alpha}$. The Weyl condition $\Gamma^{11}\Theta=\Theta$ implies
\begin{empheq}{align}
    \gamma_5\Theta^{\alpha_1\alpha_2\alpha_3}&=\alpha_1\alpha_2\alpha_3\Theta^{\alpha_1\alpha_2\alpha_3}.
\end{empheq}
The Majorana condition $\Theta^*=B_{(9,1)+}\Theta$ translates to
\begin{empheq}{align}
    \Theta^{\alpha_1\alpha_2\alpha_3*}&=-iB_{(3,1)+}\Theta^{\alpha_1-\alpha_2\alpha_3}.
\end{empheq}
This last relation means that the eight complex spinors $\Theta^{\alpha_1\alpha_2\alpha_3}$ are not all independent. We choose the four spinors with $\alpha_1\alpha_2\alpha_3=1$ as independent. These are left-handed according to the Weyl condition. Moreover, the Majorana condition implies
\begin{empheq}{align}
    \overline{\Theta}\hat{\slashed{\nabla}}\Theta&=\sum_{\alpha_1\alpha_2\alpha_3=\pm1}\overline{\Theta}^{\alpha_1\alpha_2\alpha_3}\hat{\slashed{\nabla}}\Theta^{\alpha_1\alpha_2\alpha_3}
    \\\nonumber
    &=2\sum_{\alpha_1\alpha_2\alpha_3=1}\overline{\Theta}^{\alpha_1\alpha_2\alpha_3}\hat{\slashed{\nabla}}\Theta^{\alpha_1\alpha_2\alpha_3}.
\end{empheq}
\section{Compactification on $S^2$}\label{Appendix: Compactification}
In this appendix we discuss the details of the compactification on $S^2$. Recall that the deformed $AdS_2\times S^2$ metric is
\begin{empheq}{align}
    \hat{ds}^2&=R^2\left(ds^2_H+d\Omega_2^2\right),
\end{empheq}
where $R=L\sinh(u_k)$. For a sphere with radius $R$ we use the properly normalized real spherical harmonics
\begin{empheq}{alignat=2}
    \hat{g}^{ij}\nabla_i\nabla_jY^{lm}&=-\frac{l(l+1)}{R^2}Y^{lm},
    &\qquad
    R^2\oint d\Omega_2\,Y^{lm}Y^{l'm'}&=\delta^{ll'}\delta^{mm'}.
\end{empheq}
\subsection{Scalars}
A scalar field on $AdS_2\times S^2$ can be expanded as
\begin{equation}
    \phi(\sigma^0,\sigma^1,\theta,\phi)=\sum_{l=0}^{\infty}\sum_{m=-l}^l\phi_{lm}(\sigma^0,\sigma^1)Y^{lm}(\theta,\phi).
\end{equation}
Then, the reduction of the action is
\begin{empheq}{align}
    \int d^4\sigma\sqrt{\hat{g}}\hat{g}^{\alpha\beta}\partial_{\alpha}\phi\partial_{\beta}\phi&=\sum_{l=0}^{\infty}\sum_{m=-l}^l\int d^2\sigma\sqrt{\hat{g}}\left(\hat{g}^{\mu\nu}\partial_{\mu}\phi_{lm}\partial_{\nu}\phi_{lm}+\frac{l(l+1)}{R^2}\left(\phi_{lm}\right)^2\right).
\end{empheq}
On the right hand side, in a slight abuse of notation, we have used $\hat{g}$ for the determinant of the $AdS_2$ factor of the metric.
\subsection{Gauge Field}
In order to expand the gauge field we define the following vector fields on $S^2$:
\begin{empheq}{align}
    Y^{lm}_i(\theta,\phi)=\frac{R}{\sqrt{l(l+1)}}\partial_iY^{lm}(\theta,\phi),
    \qquad
    \hat{Y}^{lm}_i(\theta,\phi)=\epsilon_i^{\phantom{i}j}Y^{lm}_j(\theta,\phi),
    \qquad
    l\geq1.
\end{empheq}
Here $\epsilon$ is the covariantly constant tensor. They satisfy
\begin{empheq}{alignat=4}
    \nabla^iY^{lm}_i&=-\frac{\sqrt{l(l+1)}}{R}Y^{lm},
    &\quad
    \epsilon^{ij}\partial_iY^{lm}_j&=0,
    &\quad
    \nabla^i\hat{Y}^{lm}_i&=0,
    &\quad
    \epsilon^{ij}\partial_i\hat{Y}^{lm}_j&=\frac{\sqrt{l(l+1)}}{R}Y^{lm},
\end{empheq}
as well as,
\begin{empheq}{align}
    R^2\oint d\Omega\,g^{ij}Y^{lm}_iY^{l'm'}_j&=\delta^{ll'}\delta^{mm'},
    \\\nonumber
    R^2\oint d\Omega\,g^{ij}\hat{Y}^{lm}_i\hat{Y}^{l'm'}_j&=\delta^{ll'}\delta^{mm'},
    \\\nonumber
    R^2\oint d\Omega\,g^{ij}Y^{lm}_i\hat{Y}^{l'm'}_j&=0.
\end{empheq}
Moreover, they form a complete set of vector fields on $S^2$. Thus, we can decompose a gauge field on $AdS_2\times S^2$ as
\begin{empheq}{align}
    a_{\mu}(\sigma^0,\sigma^1,\theta,\phi)&=\sum_{l=0}^{\infty}\sum_{m=-l}^la_{\mu}^{lm}(\sigma^0,\sigma^1)Y^{lm}(\theta,\phi),
    \\
    a_i(\sigma^0,\sigma^1,\theta,\phi)&=\sum_{l=1}^{\infty}\sum_{m=-l}^l\left(a_{lm}(\sigma^0,\sigma^1)\hat{Y}^{lm}_i(\theta,\phi)+b_{lm}(\sigma^0,\sigma^1)Y^{lm}_i(\theta,\phi)\right).
\end{empheq}
Notice that the expansion for the components $a_i$ starts at $l=1$.

Now, under a gauge transformation
\begin{empheq}{align}
    a'_{\alpha}=a_{\alpha}-\partial_{\alpha}\Lambda,
\end{empheq}
with
\begin{empheq}{align}
    \Lambda(\sigma^0,\sigma^1,\theta,\phi)=\sum_{l=0}^{\infty}\sum_{m=-l}^lc_{lm}(\sigma^0,\sigma^1)Y^{lm}(\theta,\phi),
\end{empheq}
we have
\begin{empheq}{alignat=2}
    a'_{\mu}&=\sum_{l=0}^{\infty}\sum_{m=-l}^l\left(a_{\mu}^{lm}-\partial_{\mu}c_{lm}\right)Y^{lm},
    \\
    a'_i&=\sum_{l=1}^{\infty}\sum_{m=-l}^l\left(a_{lm}\hat{Y}^{lm}_i+\left(b_{lm}-\frac{\sqrt{l(l+1)}}{R}c_{lm}\right)Y^{lm}_i\right).
\end{empheq}
By choosing $c_{lm}=R/\sqrt{l(l+1)}b_{lm}$ we can gauge fix $b'_{lm}=0$ and consider the following ansatz for the gauge field:
\begin{empheq}{align}
    a_{\mu}(\sigma^0,\sigma^1,\theta,\phi)&=\sum_{l=0}^{\infty}\sum_{m=-l}^la_{\mu}^{lm}(\sigma^0,\sigma^1)Y^{lm}(\theta,\phi),
    \\
    a_i(\sigma^0,\sigma^1,\theta,\phi)&=\sum_{l=1}^{\infty}\sum_{m=-l}^la_{lm}(\sigma^0,\sigma^1)\hat{Y}^{lm}_i(\theta,\phi).
\end{empheq}
The residual gauge symmetry is
\begin{empheq}{align}
    a^{00'}_{\mu}=a^{00}_{\mu}-\partial_{\mu}c_{00}.
\end{empheq}

Substituting this in the action we get
\begin{empheq}{align}
    \int d^4\sigma\sqrt{\hat{g}}\hat{g}^{\alpha\gamma}\hat{g}^{\beta\delta}F_{\alpha\beta}F_{\gamma\delta}&=\sum_{l=0}^{\infty}\sum_{m=-l}^l\int d^2\sigma\sqrt{\hat{g}}\left[\hat{g}^{\mu\nu}\hat{g}^{\rho\sigma}f^{lm}_{\mu\rho}f^{lm}_{\nu\sigma}+\frac{2l(l+1)}{R^2}\hat{g}^{\mu\nu}a^{lm}_{\mu}a^{lm}_{\nu}\right.
    \\\nonumber
    &\left.+2\hat{g}^{\mu\nu}\partial_{\mu}a_{lm}\partial_{\nu}a_{lm}+\frac{2l(l+1)}{R^2}\left(a_{lm}\right)^2\right],
\end{empheq}
where
\begin{empheq}{align}
    f^{lm}_{\mu\nu}&=\partial_{\mu}a^{lm}_{\nu}-\partial_{\nu}a^{lm}_{\mu}.
\end{empheq}
\subsection{Fermions}
For the expansion of fermionic fields we introduce the eigenspinors of the Dirac operator on the two-sphere which satisfy (see \cite{deBoer:1998ip,Michelson:1999kn,Corley:1999uz})
\begin{empheq}{alignat=2}
    \hat{\gamma}^i\hat{\nabla}_i\chi^{\pm}_{lm}&=\pm i\mu_l\chi^{\pm}_{lm},
    &\qquad
    \mu_l&=\frac{\left(l+\frac{1}{2}\right)}{R},
\end{empheq}
with $l=\frac{1}{2},\frac{3}{2},\ldots$ and $m=-l,-l+1,\ldots,l-1,l$. They form a complete set for spinor fields on $S^2$ and satisfy the orthonormality relations
\begin{empheq}{align}
    R^2\oint d\Omega\,\chi^{s\dagger}_{lm}\chi^{s'}_{l'm'}&=\delta^{ss'}\delta_{ll'}\delta_{mm'}.
\end{empheq}
The relative phase can be chosen so that
\begin{empheq}{align}
    \gamma_{\underline{2}\underline{3}}\chi^{\pm}_{lm}&=\pm i\chi^{\mp}_{lm}.
\end{empheq}

Spinor fields on $AdS_2\times S^2$ can be decomposed as
\begin{empheq}{align}
    \psi(\sigma^0,\sigma^1,\theta,\phi)&=\sum_{l=\frac{1}{2}}^{\infty}\sum_{m=-l}^{l}\left(\psi^+_{lm}(\sigma^0,\sigma^1)\otimes\chi^+_{lm}(\theta,\phi)+\psi^-_{lm}(\sigma^0,\sigma^1)\otimes\chi^-_{lm}(\theta,\phi)\right).
\end{empheq}
Using the gamma matrix representation
\begin{empheq}{alignat=2}
    \Gamma_{\underline{\mu}}&=\gamma_{\underline{\mu}}\otimes\mathds{1},
    &\qquad
    \Gamma_{\underline{i}}&=\gamma\otimes\gamma_{\underline{i}},
\end{empheq}
where $\gamma=\gamma_{\underline{0}\underline{1}}$, we find
\begin{empheq}{align}
    \hat{\Gamma}^{\alpha}\hat{\nabla}_{\alpha}\psi&=\sum_{l=\frac{1}{2}}^{\infty}\sum_{m=-l}^{l}\left(\hat{\gamma}^{\mu}\hat{\nabla}_{\mu}+i\mu_l\gamma\right)\psi^+_{lm}\otimes\chi^+_{lm}+\sum_{l=\frac{1}{2}}^{\infty}\sum_{m=-l}^{l}\left(\hat{\gamma}^{\mu}\hat{\nabla}_{\mu}-i\mu_l\gamma\right)\psi^-_{lm}\otimes\chi^-_{lm}.
\end{empheq}
The fermionic action then reads
\begin{empheq}{align}
    \int d^4\sigma\sqrt{\hat{g}}\bar{\psi}\hat{\Gamma}^{\alpha}\hat{\nabla}_{\alpha}\psi&=\sum_{l=\frac{1}{2}}^{\infty}\sum_{m=-l}^{l}\int d^2\sigma\sqrt{\hat{g}}\bar{\psi}^+_{lm}\left(\hat{\gamma}^{\mu}\hat{\nabla}_{\mu}+i\mu_l\gamma\right)\psi^+_{lm}
    \\\nonumber
    &+\sum_{l=\frac{1}{2}}^{\infty}\sum_{m=-l}^{l}\int d^2\sigma\sqrt{\hat{g}}\bar{\psi}^-_{lm}\left(\hat{\gamma}^{\mu}\hat{\nabla}_{\mu}-i\mu_l\gamma\right)\psi^-_{lm}.
\end{empheq}

In this paper we deal with four-dimensional Weyl spinors. Using the above decomposition, the Weyl condition implies
\begin{empheq}{align}
    \gamma\psi^{\pm}_{lm}&=\mp i\psi^{\mp}_{lm}.
\end{empheq}
With this we can eliminate $\psi^-_{lm}$ in favor of $\psi^+_{lm}$ or vise-versa. Dropping the superscript we get
\begin{empheq}{align}
    \int d^4\sigma\sqrt{\hat{g}}\bar{\psi}\hat{\gamma}^{\alpha}\hat{\nabla}_{\alpha}\psi&=2\sum_{l=\frac{1}{2}}^{\infty}\sum_{m=-l}^{l}\int d^2\sigma\sqrt{\hat{g}}\,\bar{\psi}_{lm}\left(\hat{\gamma}^{\mu}\hat{\nabla}_{\mu}+i\mu_l\gamma\right)\psi_{lm},
\end{empheq}
where $\psi_{lm}$ are unconstrained 2d Dirac spinors.
\section{The $OSp\left(4^*|4\right)$ Algebra}\label{Appendix: OSp(4*|4) Algebra}
In this section we briefly review the representations of $OSp(4^*|4)$ relevant in this paper. We closely follow \cite{Gunaydin1991}.

The $OSp\left(2m^*|2n\right)\supset SO(2m^*)\times USp(2n)$ algebra has a Jordan structure with respect to its maximum subalgebra $g^{0}=U(m|n)\supset U(m)\times U(n)$, this is,
\begin{empheq}{alignat=2}
    g&=g^{-1}\oplus g^0\oplus g^{+1},
    &\qquad
    [g^m,g^n]&\subseteq g^{m+n}
\end{empheq}
with $g^m=0$ for $|m|>1$. This decomposition is at the heart of the representation theory of this superalgebra. Denoting $A_{AB}\in g^{-1}$, $M^A_{\phantom{A}B}\in g^{0}$, $A^{AB}\in g^{+1}$, the commutation relations of $OSp\left(2m^*|2n\right)$ read
\begin{empheq}{align}
    [M^A_{\phantom{A}B},M^C_{\phantom{C}D}]&=\delta^C_{\phantom{C}B}M^A_{\phantom{A}D}-\left(-1\right)^{\left(\textrm{deg}\,A+\textrm{deg}\,B\right)\left(\textrm{deg}\,C+\textrm{deg}\,D\right)}\delta^A_{\phantom{A}D}M^C_{\phantom{C}B},
    \\
    [M^A_{\phantom{A}B},A_{CD}]&=-\delta^A_{\phantom{A}C}A_{BD}-\delta^A_{\phantom{A}D}A_{CB},
    \\
    [M^A_{\phantom{A}B},A^{CD}]&=\delta^C_{\phantom{C}B}A^{AD}+\delta^D_{\phantom{D}B}A_{CA},
    \\
    [A_{AB},A^{CD}]&=\delta^C_{\phantom{C}B}M^D_{\phantom{D}A}+\textrm{permutations}.
\end{empheq}
The index $A=(i,\mu)$, $i=1,\ldots,m$, $\mu=1,\ldots,n$ is in the fundamental of $U(m|n)$. Also, $\textrm{deg}(i)=-\textrm{deg}(\mu)=1$.

This algebra can be realized by introducing $f$ pairs of super oscillators $\xi_A(r)$ and $\eta_A(r)$, $r=1,\ldots,f$,
\begin{empheq}{alignat=2}
    \xi_A(r)&=\left(
                \begin{array}{c}
                  a_i(r) \\
                  \alpha_{\mu}(r) \\
                \end{array}
              \right),
    &\qquad
    \eta_A(r)&=\left(
                 \begin{array}{c}
                   b_i(r) \\
                   \beta_{\mu}(r) \\
                 \end{array}
               \right),
\end{empheq}
where $a$ and $b$ are bosonic and $\alpha$ and $\beta$ are fermionic. The $OSp\left(2m^*|2n\right)$ generators are then
\begin{empheq}{align}
    A_{AB}&=\xi_A\cdot\eta_B-\eta_A\cdot\xi_B,
    \\
    A^{AB}&=\eta^B\cdot\xi^A-\xi^B\cdot\eta^A,
    \\
    M^A_{\phantom{A}B}&=\xi^A\cdot\xi_B+(-1)^{\left(\textrm{deg}\,A\right)\left(\textrm{deg}\,B\right)}\eta_B\cdot\eta^A.
\end{empheq}
The dot product means sum over $r$.

In the case of $m=n=2$ the bosonic bilinears
\begin{empheq}{alignat=3}
    A_{ij}&=a_i\cdot b_j-b_i\cdot a_j,
    &\quad
    A^{ij}&=b^j\cdot a^i-a^j\cdot b^i,
    &\quad
    M^i_{\phantom{i}j}&=a^i\cdot a_j+b_j\cdot b^i,
\end{empheq}
generate $SO(4^*)\simeq SL(2,\mathds{R})\times SU(2)$. Indeed, defining
\begin{empheq}{alignat=4}
    B^-&=\frac{1}{2}\epsilon^{ij}A_{ij},
    &\quad
    B^+&=\frac{1}{2}\epsilon_{ij}A^{ij},
    &\quad
    B^0&=\frac{1}{2}M^i_{\phantom{i}i},
    &\quad
    I^i_{\phantom{i}j}&=M^i_{\phantom{i}j}-\frac{1}{2}\delta^i_{\phantom{i}j}M^k_{\phantom{k}k}.
\end{empheq}
it is easy to see that $B$ generate $SL(2,\mathds{R})$ while $I$ generate $SU(2)$. The fermionic bilinears
\begin{empheq}{alignat=3}
    A_{\mu\nu}&=\alpha_{\mu}\cdot\beta_{\nu}-\beta_{\mu}\cdot\alpha_{\nu},
    &\quad
    A^{\mu\nu}&=\beta^{\nu}\cdot\alpha^{\mu}-\alpha^{\nu}\cdot\beta^{\mu},
    &\quad
    M^{\mu}_{\phantom{\mu}\nu}&=\alpha^{\mu}\cdot\alpha_{\nu}+\beta_{\nu}\cdot\beta^{\mu},
\end{empheq}
span $USp(4)\simeq SO(5)$.

Representations of $OSp(4^*|4)$ are formed by taking a state $|\Omega\rangle$ that transforms irreducibly under $U(m|n)$ and is annihilated by $g^{-1}$,
\begin{empheq}{align}
    A_{AB}|\Omega\rangle&=0,
\end{empheq}
and acting on it with $A^{AB}$. Such a state can, in turn, be built from the oscillator vacuum $\left|0\right\rangle$ by acting with $\xi^A=\xi^{\dagger}_A$ and $\eta^A=\eta^{\dagger}_A$. Thus, $|\Omega\rangle$ is characterized by a Young tableau of $U(m|n)$. In general,
\begin{empheq}{align}
    B^0|0\rangle&=f|0\rangle
    \\
    I^i_{\phantom{i}j}|0\rangle&=0
\end{empheq}
so $|0\rangle$ has $SL(2,\mathds{R})\times SU(2)$ quantum numbers $(h,l)=(f,0)$.

Let us work out the ultrashort multiplet obtained by starting with $|0\rangle$ for $f=1$. In this case, $|0\rangle$ has quantum numbers $(h,l)=(1,0)$. We have the following $SO^*(4)\times USp(4)$ lowest weight states in the representation:
\begin{empheq}{align}
    |0\rangle,\,A^{i\mu}|0>,\,A^{i\mu}A^{j\nu}|0>.
\end{empheq}
The even generators $A^{ij}$ (or equivalently $B^+$) and $A^{\mu\nu}$ act within a given irrep of $SO^*(4)$ and $USp(4)$, respectively. We can then compute the $SL(2,\mathds{R})\times SU(2)\times SO(5)$ quantum numbers of the states above. We find,
\begin{empheq}{align}
    \mathbf{0}&=\left(1,0,\mathbf{5}\right)\oplus\left(3/2,1/2,\mathbf{4}\right)\oplus\left(2,1,\mathbf{1}\right)
\end{empheq}

More general doubleton ($f=1$) representations of $OSp(4^*|4)$ have $SL(2,\mathds{R})\times SU(2)\times SO(5)$ content,
\begin{empheq}{align}
    \mathbf{j}&=\left(j+1,j,\mathbf{5}\right)\oplus\left(j+3/2,j+1/2,\mathbf{4}\right)\oplus\left(j+2,j+1,\mathbf{1}\right)
    \\\nonumber
    &\oplus\left(j+1/2,j-1/2,\mathbf{4}\right)\oplus\left(j+1,j,\mathbf{1}\right)\oplus\left(j,j-1,\mathbf{1}\right),
\end{empheq}
for $j>1/2$ and
\begin{empheq}{align}
    \mathbf{\frac{1}{2}}&=\left(3/2,1/2,\mathbf{5}\right)\oplus\left(2,1,\mathbf{4}\right)\oplus\left(5/2,3/2,\mathbf{1}\right)\oplus\left(1,0,\mathbf{4}\right)\oplus\left(3/2,1/2,\mathbf{1}\right),
\end{empheq}
when $j=1/2$. These are obtained by starting with the vacuum $\xi^{A_1}\xi^{A_2}\cdots\xi^{A_{2j}}|0\rangle$, which has $SL(2,\mathds{R})\times SU(2)$ quantum numbers $\left(j+1,j\right)$.
\bibliographystyle{JHEP}
\bibliography{WLoops-bib}

\providecommand{\href}[2]{#2}\begingroup\raggedright\begin{thebibliography}{10}

\bibitem{Maldacena:1998im}
J.~M. Maldacena, {\it {Wilson loops in large N field theories}},  {\em Phys.
  Rev. Lett.} {\bf 80} (1998) 4859--4862,
  [\href{http://arxiv.org/abs/hep-th/9803002}{{\tt hep-th/9803002}}].

\bibitem{Rey:1998ik}
S.-J. Rey and J.-T. Yee, {\it {Macroscopic strings as heavy quarks in large N
  gauge theory and anti-de Sitter supergravity}},  {\em Eur. Phys. J.} {\bf
  C22} (2001) 379--394, [\href{http://arxiv.org/abs/hep-th/9803001}{{\tt
  hep-th/9803001}}].

\bibitem{Drukker:1999zq}
N.~Drukker, D.~J. Gross, and H.~Ooguri, {\it {Wilson loops and minimal
  surfaces}},  {\em Phys. Rev.} {\bf D60} (1999) 125006,
  [\href{http://arxiv.org/abs/hep-th/9904191}{{\tt hep-th/9904191}}].

\bibitem{Drukker:2000rr}
N.~Drukker and D.~J. Gross, {\it {An exact prediction of N = 4 SUSYM theory for
  string theory}},  {\em J. Math. Phys.} {\bf 42} (2001) 2896--2914,
  [\href{http://arxiv.org/abs/hep-th/0010274}{{\tt hep-th/0010274}}].

\bibitem{Erickson:2000af}
J.~K. Erickson, G.~W. Semenoff, and K.~Zarembo, {\it {Wilson loops in N = 4
  supersymmetric Yang-Mills theory}},  {\em Nucl. Phys.} {\bf B582} (2000)
  155--175, [\href{http://arxiv.org/abs/hep-th/0003055}{{\tt hep-th/0003055}}].

\bibitem{Pestun:2007rz}
V.~Pestun, {\it {Localization of gauge theory on a four-sphere and
  supersymmetric Wilson loops}},  \href{http://arxiv.org/abs/0712.2824}{{\tt
  arXiv:0712.2824}}.

\bibitem{Drukker:2005kx}
N.~Drukker and B.~Fiol, {\it {All-genus calculation of Wilson loops using
  D-branes}},  {\em JHEP} {\bf 02} (2005) 010,
  [\href{http://arxiv.org/abs/hep-th/0501109}{{\tt hep-th/0501109}}].

\bibitem{Gomis:2006im}
J.~Gomis and F.~Passerini, {\it {Wilson loops as D3-branes}},  {\em JHEP} {\bf
  01} (2007) 097, [\href{http://arxiv.org/abs/hep-th/0612022}{{\tt
  hep-th/0612022}}].

\bibitem{Gomis:2006sb}
J.~Gomis and F.~Passerini, {\it {Holographic Wilson loops}},  {\em JHEP} {\bf
  08} (2006) 074, [\href{http://arxiv.org/abs/hep-th/0604007}{{\tt
  hep-th/0604007}}].

\bibitem{Yamaguchi:2006tq}
S.~Yamaguchi, {\it {Wilson loops of anti-symmetric representation and D5-
  branes}},  {\em JHEP} {\bf 05} (2006) 037,
  [\href{http://arxiv.org/abs/hep-th/0603208}{{\tt hep-th/0603208}}].

\bibitem{Hartnoll:2006is}
S.~A. Hartnoll and S.~P. Kumar, {\it {Higher rank Wilson loops from a matrix
  model}},  {\em JHEP} {\bf 08} (2006) 026,
  [\href{http://arxiv.org/abs/hep-th/0605027}{{\tt hep-th/0605027}}].

\bibitem{Yamaguchi:2007ps}
S.~Yamaguchi, {\it {Semi-classical open string corrections and symmetric Wilson
  loops}},  {\em JHEP} {\bf 06} (2007) 073,
  [\href{http://arxiv.org/abs/hep-th/0701052}{{\tt hep-th/0701052}}].

\bibitem{Giombi:2006de}
S.~Giombi, R.~Ricci, and D.~Trancanelli, {\it {Operator product expansion of
  higher rank Wilson loops from D-branes and matrix models}},  {\em JHEP} {\bf
  10} (2006) 045, [\href{http://arxiv.org/abs/hep-th/0608077}{{\tt
  hep-th/0608077}}].

\bibitem{Yamaguchi:2006te}
S.~Yamaguchi, {\it {Bubbling geometries for half BPS Wilson lines}},  {\em Int.
  J. Mod. Phys.} {\bf A22} (2007) 1353--1374,
  [\href{http://arxiv.org/abs/hep-th/0601089}{{\tt hep-th/0601089}}].

\bibitem{Lunin:2006xr}
O.~Lunin, {\it {On gravitational description of Wilson lines}},  {\em JHEP}
  {\bf 06} (2006) 026, [\href{http://arxiv.org/abs/hep-th/0604133}{{\tt
  hep-th/0604133}}].

\bibitem{D'Hoker:2007fq}
E.~D'Hoker, J.~Estes, and M.~Gutperle, {\it {Gravity duals of half-BPS Wilson
  loops}},  {\em JHEP} {\bf 06} (2007) 063,
  [\href{http://arxiv.org/abs/0705.1004}{{\tt arXiv:0705.1004}}].

\bibitem{Okuda:2008px}
T.~Okuda and D.~Trancanelli, {\it {Spectral curves, emergent geometry, and
  bubbling solutions for Wilson loops}},  {\em JHEP} {\bf 09} (2008) 050,
  [\href{http://arxiv.org/abs/0806.4191}{{\tt arXiv:0806.4191}}].

\bibitem{Gomis:2008qa}
J.~Gomis, S.~Matsuura, T.~Okuda, and D.~Trancanelli, {\it {Wilson loop
  correlators at strong coupling: from matrices to bubbling geometries}},  {\em
  JHEP} {\bf 08} (2008) 068, [\href{http://arxiv.org/abs/0807.3330}{{\tt
  arXiv:0807.3330}}].

\bibitem{Forste1999}
S.~Forste, D.~Ghoshal, and S.~Theisen, {\it {Stringy corrections to the Wilson
  loop in N = 4 super Yang-Mills theory}},  {\em JHEP} {\bf 08} (1999) 013,
  [\href{http://arxiv.org/abs/hep-th/9903042}{{\tt hep-th/9903042}}].

\bibitem{Drukker:2000ep}
N.~Drukker, D.~J. Gross, and A.~A. Tseytlin, {\it {Green-Schwarz string in
  AdS(5) x S(5): Semiclassical partition function}},  {\em JHEP} {\bf 04}
  (2000) 021, [\href{http://arxiv.org/abs/hep-th/0001204}{{\tt
  hep-th/0001204}}].

\bibitem{Sakaguchi:2007ea}
M.~Sakaguchi and K.~Yoshida, {\it {A Semiclassical String Description of Wilson
  Loop with Local Operators}},  {\em Nucl. Phys.} {\bf B798} (2008) 72--88,
  [\href{http://arxiv.org/abs/0709.4187}{{\tt arXiv:0709.4187}}].

\bibitem{Kruczenski:2008zk}
M.~Kruczenski and A.~Tirziu, {\it {Matching the circular Wilson loop with dual
  open string solution at 1-loop in strong coupling}},  {\em JHEP} {\bf 05}
  (2008) 064, [\href{http://arxiv.org/abs/0803.0315}{{\tt arXiv:0803.0315}}].

\bibitem{Berenstein:1998ij}
D.~E. Berenstein, R.~Corrado, W.~Fischler, and J.~M. Maldacena, {\it {The
  operator product expansion for Wilson loops and surfaces in the large N
  limit}},  {\em Phys. Rev.} {\bf D59} (1999) 105023,
  [\href{http://arxiv.org/abs/hep-th/9809188}{{\tt hep-th/9809188}}].

\bibitem{Marolf:2003ye}
D.~Marolf, L.~Martucci, and P.~J. Silva, {\it {Fermions, T-duality and
  effective actions for D-branes in bosonic backgrounds}},  {\em JHEP} {\bf 04}
  (2003) 051, [\href{http://arxiv.org/abs/hep-th/0303209}{{\tt
  hep-th/0303209}}].

\bibitem{Marolf:2003vf}
D.~Marolf, L.~Martucci, and P.~J. Silva, {\it {Actions and fermionic symmetries
  for D-branes in bosonic backgrounds}},  {\em JHEP} {\bf 07} (2003) 019,
  [\href{http://arxiv.org/abs/hep-th/0306066}{{\tt hep-th/0306066}}].

\bibitem{Martucci:2005rb}
L.~Martucci, J.~Rosseel, D.~Van~den Bleeken, and A.~Van~Proeyen, {\it {Dirac
  actions for D-branes on backgrounds with fluxes}},  {\em Class. Quant. Grav.}
  {\bf 22} (2005) 2745--2764, [\href{http://arxiv.org/abs/hep-th/0504041}{{\tt
  hep-th/0504041}}].

\bibitem{Martucci:2005ht}
L.~Martucci and P.~Smyth, {\it {Supersymmetric D-branes and calibrations on
  general N = 1 backgrounds}},  {\em JHEP} {\bf 11} (2005) 048,
  [\href{http://arxiv.org/abs/hep-th/0507099}{{\tt hep-th/0507099}}].

\bibitem{Martucci:2006ij}
L.~Martucci, {\it {D-branes on general N = 1 backgrounds: Superpotentials and
  D-terms}},  {\em JHEP} {\bf 06} (2006) 033,
  [\href{http://arxiv.org/abs/hep-th/0602129}{{\tt hep-th/0602129}}].

\bibitem{DeWolfe:2001pq}
O.~DeWolfe, D.~Z. Freedman, and H.~Ooguri, {\it {Holography and defect
  conformal field theories}},  {\em Phys. Rev.} {\bf D66} (2002) 025009,
  [\href{http://arxiv.org/abs/hep-th/0111135}{{\tt hep-th/0111135}}].

\bibitem{Kruczenski:2003be}
M.~Kruczenski, D.~Mateos, R.~C. Myers, and D.~J. Winters, {\it {Meson
  spectroscopy in AdS/CFT with flavour}},  {\em JHEP} {\bf 07} (2003) 049,
  [\href{http://arxiv.org/abs/hep-th/0304032}{{\tt hep-th/0304032}}].

\bibitem{Karch:2005ms}
A.~Karch, A.~O'Bannon, and K.~Skenderis, {\it {Holographic renormalization of
  probe D-branes in AdS/CFT}},  {\em JHEP} {\bf 04} (2006) 015,
  [\href{http://arxiv.org/abs/hep-th/0512125}{{\tt hep-th/0512125}}].

\bibitem{Arean:2006pk}
D.~Arean and A.~V. Ramallo, {\it {Open string modes at brane intersections}},
  {\em JHEP} {\bf 04} (2006) 037,
  [\href{http://arxiv.org/abs/hep-th/0602174}{{\tt hep-th/0602174}}].

\bibitem{Kirsch:2006he}
I.~Kirsch, {\it {Spectroscopy of fermionic operators in AdS/CFT}},  {\em JHEP}
  {\bf 09} (2006) 052, [\href{http://arxiv.org/abs/hep-th/0607205}{{\tt
  hep-th/0607205}}].

\bibitem{PandoZayas:2008hw}
L.~A. Pando~Zayas, V.~G.~J. Rodgers, and K.~Stiffler, {\it {Luscher Term for
  k-string Potential from Holographic One Loop Corrections}},  {\em JHEP} {\bf
  12} (2008) 036, [\href{http://arxiv.org/abs/0809.4119}{{\tt
  arXiv:0809.4119}}].

\bibitem{Doran:2009pp}
C.~A. Doran, L.~A. Pando~Zayas, V.~G.~J. Rodgers, and K.~Stiffler, {\it
  {Tensions and Luscher Terms for (2+1)-dimensional k-strings from Holographic
  Models}},  {\em JHEP} {\bf 11} (2009) 064,
  [\href{http://arxiv.org/abs/0907.1331}{{\tt arXiv:0907.1331}}].

\bibitem{Stiffler:2009ma}
K.~Stiffler, {\it {Mesons From String Theory}},
  \href{http://arxiv.org/abs/0909.5681}{{\tt arXiv:0909.5681}}.

\bibitem{Stiffler:2010pz}
K.~Stiffler, {\it {A Walk Through Superstring Theory With an Application to
  Yang-Mills Theory: K-strings and D-branes as Gauge/Gravity Dual Objects}},
  \href{http://arxiv.org/abs/1012.0021}{{\tt arXiv:1012.0021}}.

\bibitem{Nahm:1977tg}
W.~Nahm, {\it {Supersymmetries and their representations}},  {\em Nucl. Phys.}
  {\bf B135} (1978) 149.

\bibitem{Gunaydin:1986fe}
M.~Gunaydin, G.~Sierra, and P.~K. Townsend, {\it {The unitary supermultiplets
  of d = 3 Anti-de-Sitter and d = 2 conformal superalgebras}},  {\em Nucl.
  Phys.} {\bf B274} (1986) 429.

\bibitem{deBoer:1998ip}
J.~de~Boer, {\it {Six-dimensional supergravity on S**3 x AdS(3) and 2d
  conformal field theory}},  {\em Nucl. Phys.} {\bf B548} (1999) 139--166,
  [\href{http://arxiv.org/abs/hep-th/9806104}{{\tt hep-th/9806104}}].

\bibitem{Michelson:1999kn}
J.~Michelson and M.~Spradlin, {\it {Supergravity spectrum on AdS(2) x S(2)}},
  {\em JHEP} {\bf 09} (1999) 029,
  [\href{http://arxiv.org/abs/hep-th/9906056}{{\tt hep-th/9906056}}].

\bibitem{Corley:1999uz}
S.~Corley, {\it {Mass spectrum of N = 8 supergravity on AdS(2) x S(2)}},  {\em
  JHEP} {\bf 09} (1999) 001, [\href{http://arxiv.org/abs/hep-th/9906102}{{\tt
  hep-th/9906102}}].

\bibitem{Lee:1999yu}
J.~Lee and S.~Lee, {\it {Mass Spectrum of D=11 Supergravity on AdS2 x S2 x
  T7}},  {\em Nucl. Phys.} {\bf B563} (1999) 125--149,
  [\href{http://arxiv.org/abs/hep-th/9906105}{{\tt hep-th/9906105}}].

\bibitem{Gunaydin1991}
M.~Gunaydin and R.~J. Scalise, {\it {Unitary lowest weight representations of
  the noncompact supergroup OSp(2m*/2n)}},  {\em J. Math. Phys.} {\bf 32}
  (1991) 599--606.

\bibitem{Camino:2001at}
J.~M. Camino, A.~Paredes, and A.~V. Ramallo, {\it {Stable wrapped branes}},
  {\em JHEP} {\bf 05} (2001) 011,
  [\href{http://arxiv.org/abs/hep-th/0104082}{{\tt hep-th/0104082}}].

\bibitem{Drukker:2006zk}
N.~Drukker, S.~Giombi, R.~Ricci, and D.~Trancanelli, {\it {On the D3-brane
  description of some 1/4 BPS Wilson loops}},  {\em JHEP} {\bf 04} (2007) 008,
  [\href{http://arxiv.org/abs/hep-th/0612168}{{\tt hep-th/0612168}}].

\bibitem{Zarembo:1999bu}
K.~Zarembo, {\it {Wilson loop correlator in the AdS/CFT correspondence}},  {\em
  Phys. Lett.} {\bf B459} (1999) 527--534,
  [\href{http://arxiv.org/abs/hep-th/9904149}{{\tt hep-th/9904149}}].

\bibitem{Burrington:2010yb}
B.~A. Burrington and L.~A. Pando~Zayas, {\it {Phase transitions in Wilson loop
  correlator from integrability in global AdS}},
  \href{http://arxiv.org/abs/1012.1525}{{\tt arXiv:1012.1525}}.

\bibitem{Sachdev:2010uj}
S.~Sachdev, {\it {Strange metals and the AdS/CFT correspondence}},  {\em J.
  Stat. Mech.} {\bf 1011} (2011) P11022,
  [\href{http://arxiv.org/abs/1010.0682}{{\tt arXiv:1010.0682}}].

\bibitem{Mueck:2010ja}
W.~Mueck, {\it {The Polyakov Loop of Anti-symmetric Representations as a
  Quantum Impurity Model}},  \href{http://arxiv.org/abs/1012.1973}{{\tt
  arXiv:1012.1973}}.

\end{thebibliography}\endgroup
\end{document}